# Electrohydrodynamic coupling and stochastic branching in a miniaturized ns-pulsed He plasma jet

Y. Agha[1], K. Giotis[1,2], D. Stefas[1], N. Fagnon[1], G. P. Vafakos[3], V. Vavourakis[3,4], H. Höft[5], P. Svarnas[2], G. Lombardi[1], K. Gazeli[1,*]

[1]*Laboratoire des Sciences des Procédés et des Matériaux (LSPM – CNRS), Université Sorbonne Paris Nord, Villetaneuse, UPR 3407, F-93430, France*
[2]*Plasma Technology Group, High Voltage Laboratory, Electrical and Computer Engineering Department, University of Patras, 26504 Rion–Patras, Greece*
[3]*Mechanical and Manufacturing Department, University of Cyprus, Nicosia, 2109, Cyprus*
[4]*Medical Physics and Biomedical Engineering Department, University College London, London, WC1E 6BT, United Kingdom*
[5]*Leibniz Institute for Plasma Science and Technology (INP), Felix-Hausdorff-Straße 2, 17489 Greifswald, Germany*

**Corresponding author email: kristaq.gazeli@lspm.cnrs.fr*

## Abstract

This study focuses on the complex coupling between discharge and flow properties in a ns-pulsed He micrometer scale atmospheric pressure plasma jet (μAPPJ). This is investigated by integrating electrical measurements, schlieren photography, ICCD imaging, and space-resolved Optical Emission Spectroscopy (OES) with Computational Fluid Dynamics (CFD) simulations. In the flow rate range $Q_V$=0.1–1 slm, a critical threshold emerges at 0.3 slm, where the discharge consumes the highest energy overall, achieving maximum propagation length and remarkable collimation. Below 0.3 slm, insufficient momentum renders the jet susceptible to buoyancy and air entrainment, leading to shorter effluents, while higher flow rates enhance shear layer instabilities. CFD simulations reproduce the schlieren flow profiles to quantify the axial helium mass fraction ($Y_{He}$) confirming a stable helium-rich core at 0.3 slm ($Y_{He}$=90%), not seen in other flow rates. Furthermore, lower and higher flow rates promote stochastic branching which is more pronounced at $Q_V$>0.3 slm. Numerous lateral branches are clearly distinguished and quantified via single-shot ICCD imaging for the first time in a He μAPPJ. The increase of voltage amplitude ($V_P$) in the range 4–9.5 kV, amplifies their activity in the effluent tip at 0.3 slm. At $V_P$=9.5 kV, their number increases for $Q_V$<0.3 slm compared to $Q_V$=0.3 slm, while for $Q_V$>0.3 slm they occur much closer to the nozzle exit and intensify farther downstream. Time-resolved imaging reveals a distinct peak in ionization wave velocities (up to ≈600 km/s at 0.3 slm/9.5 kV) just after the nozzle exit, followed by a progressive decay which becomes more abrupt at the higher flow rates. This correlates spatially with a surge and subsequent axial drop in $N_2^+$(FNS) emission intensity, indicating Penning ionization as a key mechanism behind this acceleration. Average gas temperature estimations ($T_{Gas}$≈350 K) suggest that localised thermal expansion could contribute to the instabilities observed, possibly combined with a sudden rise in the average electrohydrodynamic force at the nozzle exit for $Q_V$≠0.3 slm. Finally, the device geometry also plays a decisive role in internal vortex formation, especially at higher flow rates, affecting effluent stability. These results provide a unique framework for optimizing μAPPJs for high-precision applications such as in analytical chemistry and surface processing.

## 1. Introduction

Atmospheric pressure plasma jets (APPJs) are versatile non-equilibrium discharges typically consisting of a thin dielectric tube and various electrode configurations [1,2]. Usually driven by sinusoidal AC or pulsed high-voltage (HV) excitations, they can readily ionize a noble gas carrier producing luminous effluents up to several centimeters in length [3–14]. While appearing continuous to the naked eye, these effluents consist of discrete ionization waves (IWs) propagating with velocities up to several hundreds of kilometers per second [1,3,4,6,14–20]. An important advantage of APPJs is the, on average, low power consumption and their ability to generate high concentrations of reactive oxygen and nitrogen species at near-ambient gas temperatures. This makes them ideal for applications in biomedicine, food science, agriculture, analytical atomic spectrometry, and material processing [2,21–27].

The performance of APPJs is fundamentally governed by the complex coupling between discharge and flow dynamics [28–35]. Notably, the gas flow rate, electrical parameters, and APPJ geometry are decisive factors in the discharge ignition and evolution of the fluid, frequently triggering a premature laminar-to-turbulent transition. Most published works refer to millimeter-scale APPJs (mAPPJs), where the luminous effluent length typically increases with the flow rate until reaching an upper threshold that, then, decreases sharply as the flow becomes transitional or turbulent [28,36]. This behavior is dictated by the noble gas mole fraction along the propagation axis. In fact, a critical noble gas concentration is required to sustain a stable and long IW trajectory, whereas excessive air entrainment quenches the discharge [28,36]. Besides, at low gas flow rates, insufficient flow momentum transfer renders the jet susceptible to buoyancy forces, resulting in curved jets that disrupt axial consistency and directionality [29,31]. Thus, finding optimal device configurations, electrical parameters and flow rates, is vital for enhancing the APPJ stability and control for the above-mentioned applications [2].

A fundamental aspect of active research in the family of mAPPJs refers to the mechanism behind plasma-induced flow modification. This generally comprises two physical phenomena: gas heating and electrohydrodynamic (EHD) forces. In the former case, localized Ohmic heating creates density gradients that trigger shear instabilities [33,37–41], while in the latter case the so-called ionic wind plays an important role due to momentum transfer from drifted ions to neutrals which accelerates the gas stream [31,32,39,42,43]. In ns-pulsed discharges, even a single voltage pulse can trigger a turbulent front that moves in the gas channel with a velocity similar to that of the gas [34,40,44,45]. Furthermore, decreasing the APPJ dielectric tube diameter results in higher electron densities, ionization wave velocities, and electric field strengths [46–51]. These changes may further trigger flow structuration.

While both discharge and gas flow dynamics have been extensively investigated in the case of mAPPJs, the physical properties of micro-scaled jets (μAPPJs), particularly their hydrodynamics, remain underreported. Featuring luminous effluent diameters well below 1 mm [46,47,52–55], μAPPJs offer significantly reduced gas consumption and improved analytical precision. These advantages over the more classic mAPPJs make them highly attractive as soft ionization sources for analytical chemistry and ambient mass spectrometry, antimicrobial applications, etc. [22,53,56–60]. However, their comprehensive characterization is often hindered by their miniaturized dimensions and fast dynamics, which push the limits of conventional diagnostics. A number of studies have combined electrical measurements with optical diagnostics (such as space-time-resolved OES, absorption spectroscopy, ICCD imaging, and TALIF) and/or numerical models to study μAPPJs [46–48,50,53–55,61–69]. These investigations have focused on atomic ground– and excited–state concentrations, electron densities, and IW velocities, providing a foundation for better understanding μAPPJ behavior. The gas flow rate affects the propagation length in μAPPJs but has a small influence in the initial development and propagation velocity of IWs inside the dielectric tube [50]. Furthermore, while higher gas flow rates can slightly reduce gas temperatures through enhanced convective cooling [53], the transition from laminar to turbulent flow at specific Reynolds numbers remains a critical limiting factor for the effluent length [36,46]. Regarding device structural factors, although reduced tube diameters inherently increase the gas velocity, the corresponding increase in IW velocity is primarily attributed to enhanced electric fields and surface charge effects [48]. Despite these insights, direct visualization of the neutral gas flow dynamics (e.g., via schlieren imaging [70]) in tandem with advanced plasma diagnostics is largely absent in the μAPPJ literature. This is an important gap since hydrodynamic instabilities can fundamentally reshape the discharge pattern in these tiny jets, thus affecting their applicability. Consequently, a holistic understanding of discharge and gas flow dynamics in μAPPJs requires a coupled approach that correlates advanced optical diagnostics with direct flow visualization methods and numerical models.

In this work, we focus on the complex electrohydrodynamic coupling in a He μAPPJ driven by ns-pulsed HV with a fixed pulse duration and repetition frequency. We reveal critical gas flow rates and HV amplitudes inducing multiple stochastic branches in the luminous effluent region. For the first time, we clearly isolate the induced branches, quantify them, and argue about the conditions governing the transition from a stable to a stochastic branched luminous effluent. To achieve this, a unique multi-modal analysis is considered. We utilize electrical measurements, single-shot and time-resolved ICCD imaging to track the discharge global emission pattern and its dynamic evolution, schlieren imaging combined with Computational Fluid Dynamics (CFD) simulations to visualize neutral gas dynamics, and space-resolved OES to map the axial chemical reactivity, gas temperature ($T_{Gas}$), electron density and axial electric field. A specific focus is placed on identifying the optimal operational regime where the effluent reaches enhanced stability due to a balance between flow momentum-driven collimation and plasma-induced perturbations. Moreover, we demonstrate how discrete discharge ionization wave fronts interact with the helium–air interface, showing both collimated and non-collimated (bush-like) behavior, while we attempt to elucidate the role of $T_{Gas}$, electrohydrodynamic force and, notably, device geometry, on the flow stability.

The remainder of the manuscript is organized as follows. **Section 2** presents the microplasma source, experimental setup and equipment used to characterize the μAPPJ. **Section 3** outlines the CFD model used, **section 4** presents the findings of this study, **section 5** discusses the results and compares them against relevant literature, and **section 6** draws the main conclusions.

## 2. Experimental setups and methods

**Fig. 1** shows the custom-made plasma device (bottom left), which is mounted on a micrometric translation stage (detail (8)). The experimental setup used to investigate the electrical and optical features of the μAPPJ is also shown. The device features a single electrode configuration, representing a distinct APPJ type [1]. It consists of a stainless-steel hollow needle (25 mm in length; Ø350 μm and Ø180 μm outer and inner diameters, respectively), with one end terminating into a hollow base (4.2 mm inner diameter; 12 mm height). The hole of the base is aligned with a through hole (Ø4.2 mm) made on a cylindrical 3D-printed PLA body (shown in dark grey in **Fig. 1**). To achieve this, a brass forked shim firmly grips the base of the needle and is tightly fixed into a mold made on top of the PLA body. A polyurethane tube, carrying the operating gas (He Alphagaz 1 from Air Liquide; 99.999% purity with certified impurities: $H_2O \leq 3$ ppm, $O_2 \leq 2$ ppm, CnHm ≤ 0.5 ppm, CO ≤ 0.5 ppm, and $CO_2 \leq 0.5$ ppm), is passed through the hole of the PLA body, and is tightly inserted into the base hole of the needle. It is then glued using an epoxy adhesive, allowing the gas to be transferred directly into the needle while preventing leakages. The needle is inserted into a borosilicate 3.3 glass microtube (32 mm length; $d_2$=600 μm inner diameter; 90% transmittance) which is also glued on the base of the needle. The helium flow rate ($Q_V$) is adjusted between 0.1 and 1 standard liters per minute (slm) using a mass flow meter (Bronkhorst MASS-VIEW®; ±2% precision). The dielectric tube can be covered by a hollow 3D-printed PLA cup (shown in light grey in **Fig. 1**), which can be firmly screwed onto the top of the first PLA body, offering additional protection.

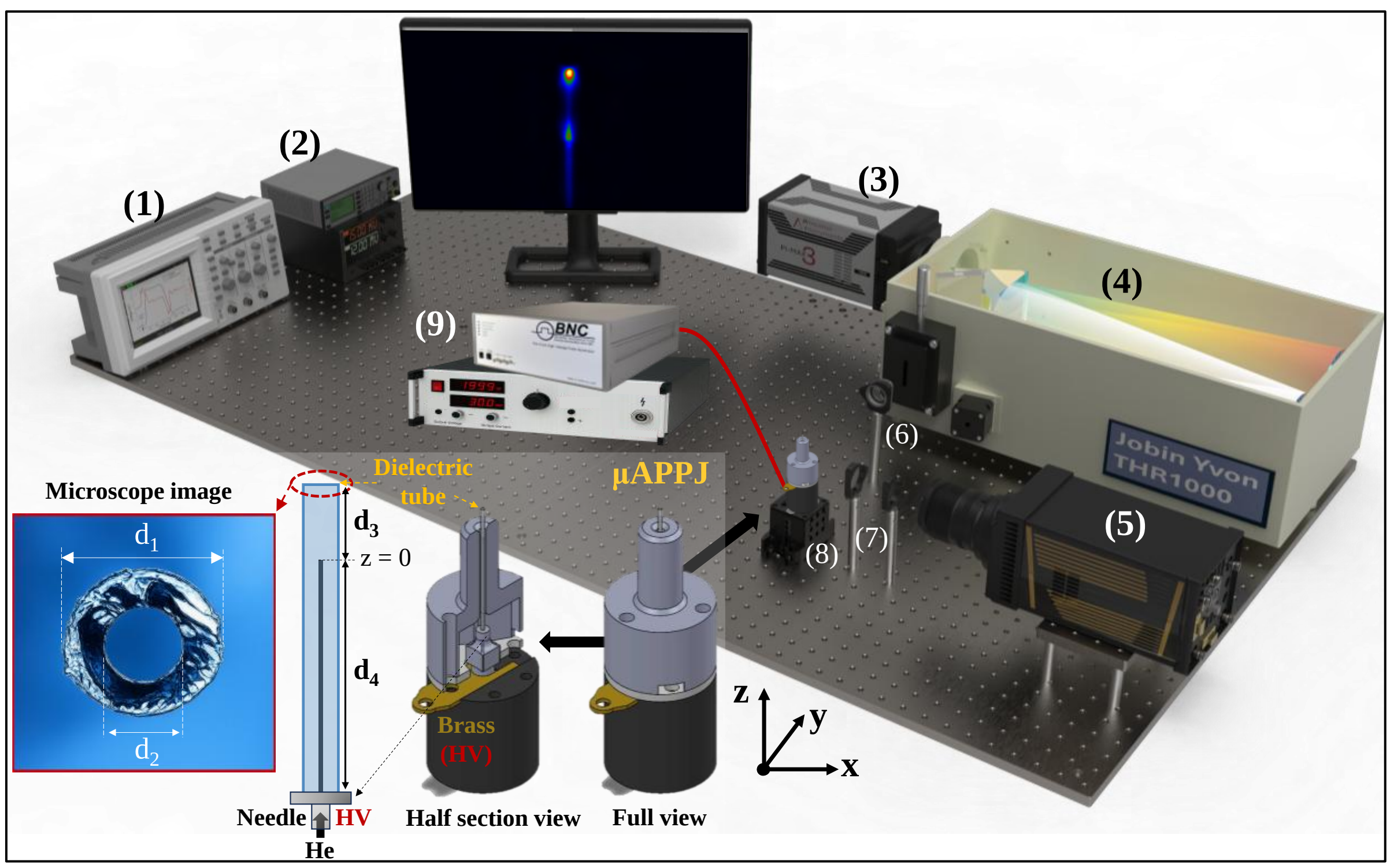

***Figure 1***. *Experimental setup used to study the electrical and optical features of the μAPPJ. Structural details of the device are shown at the bottom left, along with a magnified photo of the nozzle of the dielectric tube ($d_1$=1000 μm, $d_2$=600 μm, $d_3$=7 mm, $d_4$=25 mm). (1): digital oscilloscope, (2): delay generator and mass flow meter, (3)/(5): ICCD cameras, (4): 1 m focal length spectrometer, (6)/(7): UV lenses and linear polariser (depending on the experiment) (8): micrometric translation stage (with the μAPPJ), (9): HV DC power supply and HV pulse generator.*

To drive the μAPPJ, the brass electrode, being in contact with the base of the hollow needle, is connected to the output of a pulsed power supply (detail (9) in **Fig. 1**). This comprises a positive high-voltage (HV) DC generator (F.u.G. Elektronik GmbH HCN700-20000), an HV pulse generator (Berkeley Nucleonics Corp. PVX-4110), and a digital delay generator (Stanford Research Systems DG645; 25 ps rms jitter). This system allows for the production of rectangular HV pulses with a peak voltage ($V_P$) up to +10 kV, a repetition frequency ($f_{pulse}$) up to 10 kHz, and a minimum pulse duration ($\Delta t_{pulse}$) of 160 ns. Thus, high instantaneous power can be achieved while maintaining the average gas temperature and power consumption at low levels. The average discharge current is also reduced, minimizing the probability of arcing. In this work, $f_{pulse}$ and $\Delta t_{pulse}$ are fixed at 10 kHz and 160 ns, respectively. To monitor the applied HV waveform, a high-voltage passive probe (Tektronix P6015A, DC–75 MHz) is directly connected to the brass electrode, while the current is measured on the HV line using a current transformer (Magnelab CT-D2.5-BNC, 1200 Hz–500 MHz). To achieve this, the HV cable is passed through the current transformer using a cylindrical PVC bushing for insulation. The electrical waveforms are recorded on a wideband digital oscilloscope (Teledyne LeCroy WaveSurfer10; 1 GHz bandwidth; 10 GS/s sampling rate). All signals are synchronized between them (±1 ns estimated jitter) by considering all system delays (cables, equipment, etc.). The discharge is operated for 30 minutes before experimentation to ensure stable operation and reliable comparison between different operating conditions.

To record the fast dynamics of the total (wavelength-integrated) discharge emission, an ICCD camera (Princeton Instruments PIMAX-1K-RB-FG-P43) is used (detail (5) in **Fig. 1**). It features a gated intensifier (2 ns minimum gate width), an image sensor (Marconi CCD47-10, 1024×1024 $px^2$ with 13 μm pixel size), and a controller (ST-133; PTG as internal pulse generator). To capture sharp, focused images of the discharge inside and outside the microtube, either an UV objective lens (CERCO®) or a set of two UV fused silica plano-convex lenses (Thorlabs LA4184—f=500 mm and LA4579—f=300 mm) is used. The objective lens is directly mounted on the camera while the two UV lenses are positioned back-to-back (detail (7) in **Fig. 1**), with both the lens-to-ICCD and lens-to-μAPPJ distances set to twice the effective focal length ($f_{eff}$≈187.5 mm). This configuration records a 13.8×13.8 $mm^2$ area. As this field of view cannot capture the entire discharge length, the two-lens system is only used for cross-validation of the results obtained with the UV objective lens. The camera is triggered by the same digital delay generator used for triggering the HV pulse generator, which also allows for time-resolved emission analysis by adjusting the gate delay with respect to the rising edge of the HV pulse. Unless otherwise specified, for time-resolved studies every ICCD image is captured with the following settings: 2 ns gate, on-chip accumulation over 100 HV pulses, 2 ns step, and a gain of 230. The whole duration of the HV pulse is scanned in order to follow the discharge dynamics during the rising and falling phases of the applied voltage pulse. The reference axial position (z=0 mm) along the propagation axis (z) for performing space-resolved optical studies inside the microtube and along the effluent is set at the tip of the needle (7 mm before the exit nozzle of the dielectric tube; see **Fig. 1**). The same camera is also used to capture single shot (gate width: 300 ns, i.e., capturing the total emission over one HV pulse) and averaged (300 ns gate width and post-averaging over 500 HV pulses) images of the discharge pattern.

The velocity and maximum propagation length of ionization waves are determined through an analysis of raw ICCD images using a custom-developed Python routine. To mitigate the influence of background noise and spurious low-intensity pixels, an initial low-intensity clipping step is applied to suppress values below a predefined noise threshold. The images are then processed with a median filter for further noise reduction. To extract the spatial light distribution, the pixel intensities are integrated row-wise to generate a longitudinal intensity profile. For each resulting profile, a linear baseline is subtracted and a secondary 1D median filter is applied for additional denoising. The position of the ionization wave front is defined as the leading-edge front location where the intensity decays to 5% of the peak value. Finally, the ionization wave front velocity ($u$) is calculated from the distance $\Delta z$ traversed by the ionization wave front between consecutive frames over the time interval $\Delta t$, according to the formula $u = \Delta z / \Delta t$. This methodology is similar to that detailed in our recent work [71].

To identify the emissive species formed in the discharge, optical emission spectroscopy (OES) is carried out using a high-resolution spectrometer (Jobin Yvon THR 1000; 1 m focal length; 1200 l/mm grating blazed at 500 nm), equipped with an ICCD camera (Princeton Instruments PI-MAX 3 1024i; 12.8 μm pixel size; 15 ns minimum gate width). This spectrometer is also used for time-integrated and space-resolved measurements of the intensities of reactive molecular bands (OH(A–X), $N_2$(C–B), …) and atomic lines ($H_β$/$H_α$, He I, O I). Furthermore, it allows for (i) the estimation of the gas temperature ($T_{Gas}$) from the rotational temperature ($T_{Rot}$) of excited probe molecules via fitting of their rotational spectra [7,72,73], and from the pressure broadening of a He I line as explained in [69,72,73], (ii) the electron density through the stark broadening of $H_β$ line [74–76], and (iii) the axial electric field based on Stark polarization spectroscopy using the He I line around 492.2 nm [10–12,77,78]. The entrance slit width of the spectrometer is fixed to 40 μm while the μAPPJ is mounted on a micrometric translation stage with its axis being normal to the (vertical) orientation of the slit (i.e., the μAPPJ is rotated by 90° with respect to its orientation shown in **Fig. 1**). An UV lens (Thorlabs LA5817, 10 cm focal length) in a 2f—2f configuration (detail (6) in **Fig. 1**) is used to image the μAPPJ effluent onto the slit of the spectrometer and collect light from various discrete small zones (1 mm apart) along the discharge propagation axis. For the axial electric field measurements, a linear polariser (Thorlabs LPVISE200-A) is used. To enhance the signal-to-noise ratio (SNR), time-integrated spectra are recorded using variable ICCD settings (exposure time and on-chip accumulations) depending on the species studied.

To visualize the interaction between the gas flow and discharge, schlieren diagnostic is employed. As shown in **Fig. 2**, the setup uses a single-mirror off-axis configuration which is similar to other works [70,79–82]. This is practical in our case, as the difference in mass between helium and air is close to a factor of 7. The illumination source consists of a green LED (Thorlabs LED525E, 525 nm, 32 nm FHWM), followed by a glass diffuser (Thorlabs DG10-600-A**)** and a pinhole. The light is collected and reflected by a parabolic mirror (Ø7.6 cm; $f$≈44.5 cm; >90% reflectance for λ>400 nm). After reflection, the light beam passes through the µAPPJ and is focused onto a sharp razor blade (2$f$ configuration). The blade blocks approximately half of the light, converting ray deflections (induced by variations of the refractive index due to the gas density variation) into intensity gradients in the recorded images. Images are captured by a DSLR camera (Nikon D5300; f/5.3, ISO-100, exposure time: 20 ms corresponding to 200 HV pulses) coupled to an objective lens (Nikon AF-S NIKKOR 80-400 mm).

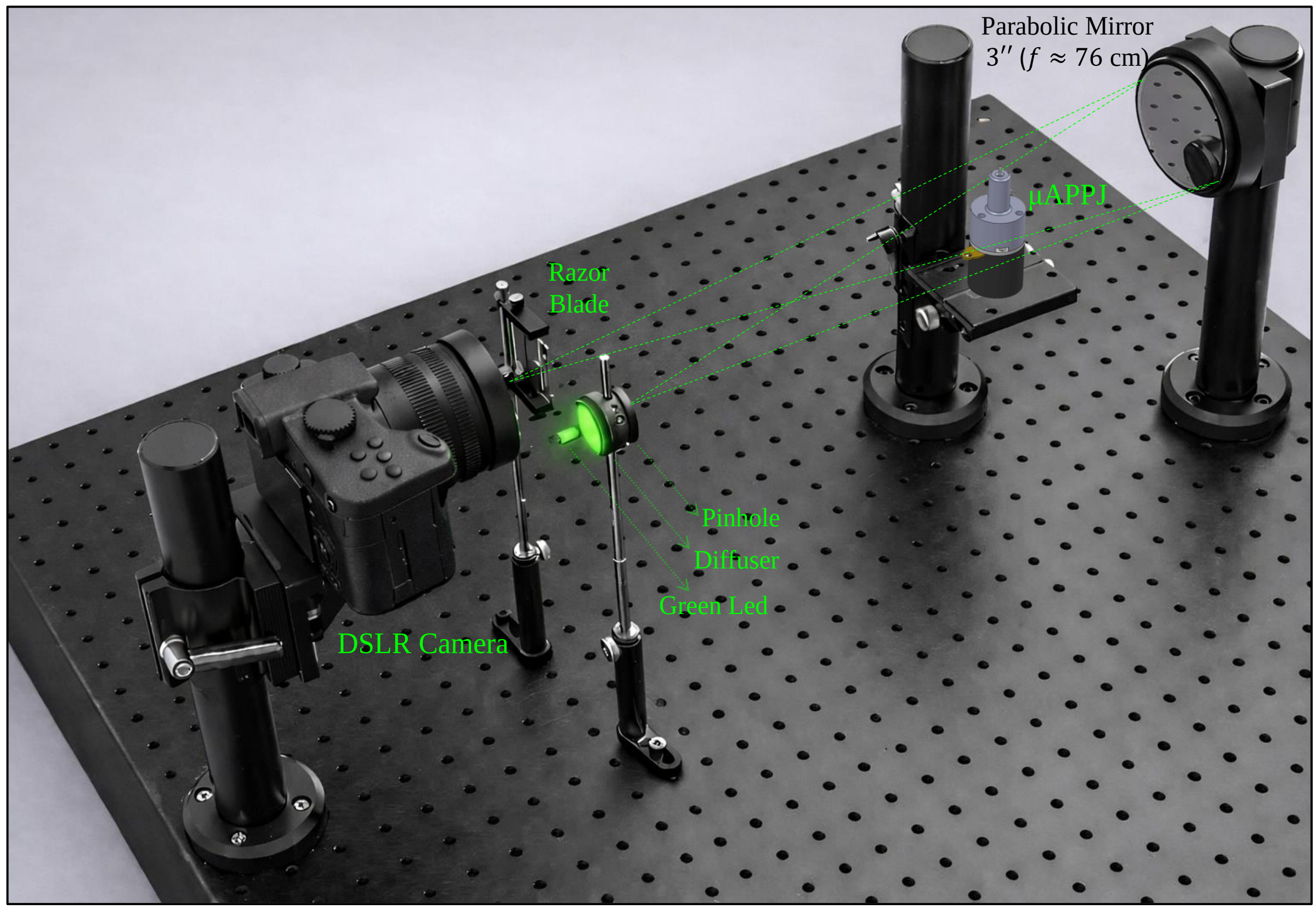

***Figure 2***. *Experimental setup used for schlieren imaging.*

## 3. Computational fluid dynamics (CFD) model

CFD simulations are carried out to further investigate the behavior of the background gas dynamics of the effluent, under both plasma "OFF" and "ON" conditions. For the simulation of gas flow, a mixture of two species is assumed: helium and artificial dry air (78% $N_2$ and 22% $O_2$). The compressible equations for the mass, momentum and energy conservation of the mixture read:

$$\frac{\partial \rho}{\partial t} + \nabla \cdot (\rho \boldsymbol{U}) = 0 \tag{1}$$

$$\frac{\partial (\rho \boldsymbol{U})}{\partial t} + \nabla \cdot \left(\rho \boldsymbol{U}\boldsymbol{U} + p\bar{\bar{I}} - \bar{\bar{\tau}}\right) - \rho \boldsymbol{g} = 0 \tag{2}$$

$$\frac{\partial (\rho E)}{\partial t} + \nabla \cdot \left(\rho \boldsymbol{U} H - \bar{\bar{\tau}} \cdot \boldsymbol{U} + \boldsymbol{q}\right) = 0 \tag{3}$$

where $p$ is the pressure, $\boldsymbol{U}$ is the gas velocity vector field, $\rho$ is the mixture density, $\bar{\bar{\tau}}$ is the viscous stress tensor, $\boldsymbol{g}$ is the gravity vector, $E = e + U^2/2$ is the sum of the internal and kinetic energy, $H = E + p/\rho$ is the total enthalpy and $\boldsymbol{q}$ is the total heat flux, if present. For the mixing of the species in the gas, the multi-component mass transport equation is used:

$$\frac{\partial}{\partial t}(\rho Y_i) + \boldsymbol{\nabla} \cdot (\rho \boldsymbol{U} Y_i) = -\boldsymbol{\nabla} \cdot \boldsymbol{j_i} \tag{4}$$

where $Y_i$ and $\boldsymbol{j_i}$ are the mass fraction and diffusive flux of species $i$, respectively. A more detailed description of eqs. (1)-(4) can be found in the literature [43,83].

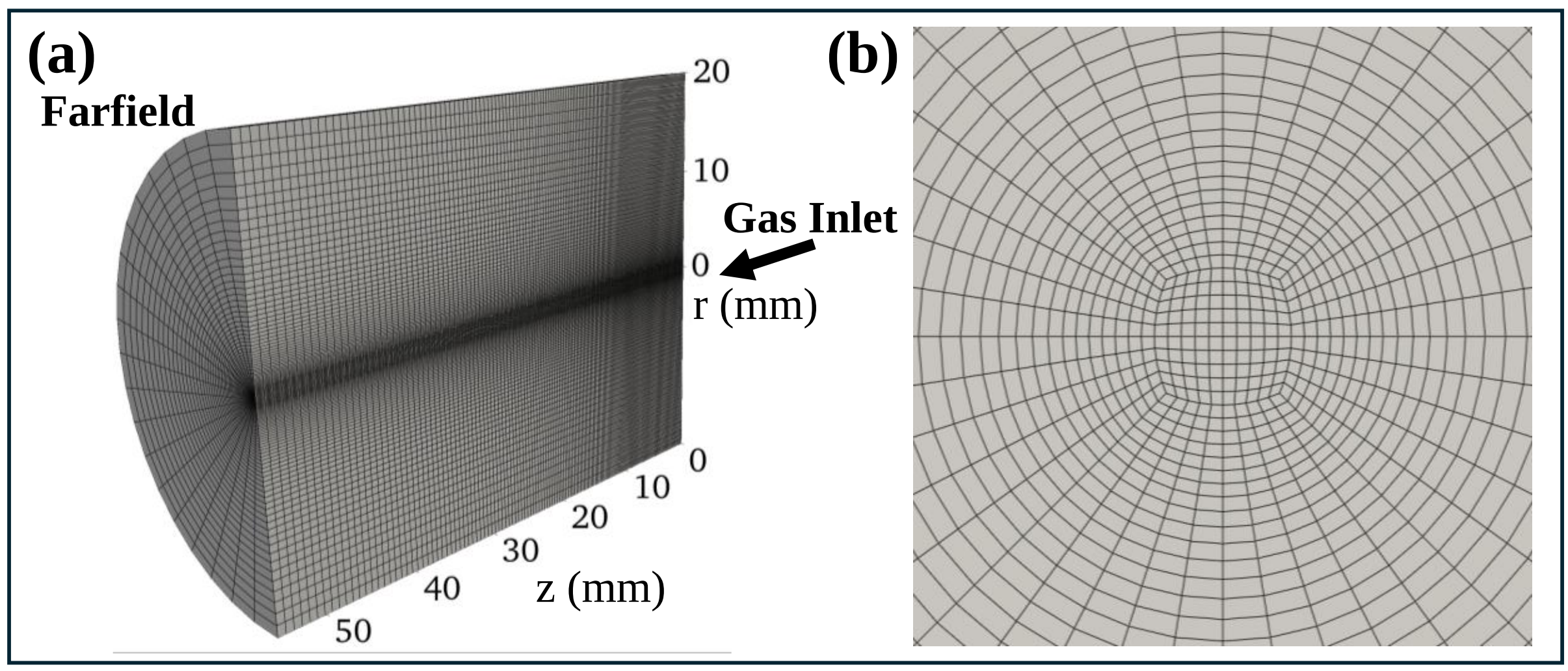

***Figure 3**. Computational mesh used for simulating the plasma reactor, depicted in (a) bird's eye, and (b) cross-sectional views.*

The governing equations of gas flow are discretized using the finite volume method and solved numerically via the open-source CFD framework OpenFOAM [84]. Turbulence is resolved using a large eddy simulation (LES) formulation with the dynamic subgrid scale model implemented within OpenFOAM. The geometry and the dimensions of the μAPPJ used in the simulations are shown in **Fig. 1**. In **Fig. 3** a schematic of the cylindrical computational domain and the corresponding mesh is presented. The farfield boundaries are positioned 20 mm radially and 55 mm axially from the exit nozzle of the capillary tube. The domain is discretized using a structured mesh of approximately 1.3 million non-uniform cells, with increasing resolution towards the location of the reactor.

Simulations are performed for four different gas flow rates (0.1, 0.3, 0.5 and 1.0 slm) and validated against the schlieren experimental data (see **sections 4**, **5**). To better describe the turbulence at the highest flow rate, special turbulent inlet boundary conditions are applied. The two different plasma "OFF" and plasma "ON" cases are introduced in the fluid dynamic equations, by imposing different temperatures at the exit of the tube. For the plasma "OFF" case the tube is set to be at room temperature, measured at 293 K at the days of the experiments, while for the plasma "ON" condition the mean gas temperature at the exit nozzle is measured to be around 350 K (see **Fig. 14**). Regarding the parameters of the model, the ideal gas assumption is applied with the molar weight of dry air and helium at 28.9 and 4.0 kg/kmol, respectively. At 293 K the densities of dry air and helium are 1.2 and 0.166 kg/m$^3$, respectively, and at 350 K they are 1.0 and 0.139 kg/m$^3$, respectively. Similarly, at 293 K the dynamic viscosities of dry air and helium are $1.81\times10^{-5}$ and $1.96\times10^{-5}$ Pa·s, respectively, and at 350 K they are $2.07\times10^{-5}$ and $2.22\times10^{-5}$ Pa·s, respectively.

## 4. Results

Using coupled electrical and optical diagnostics, the impact of the two main parameters of the system, i.e., the gas flow rate ($Q_V$) and the voltage amplitude ($V_P$), on the discharge and flow properties are investigated. **Section 4.1** interrogates the main electrical properties of μAPPJ, while the following sections focus on the averaged discharge morphology and gas flow field properties (**section 4.2**), the single-shot discharge dynamics, effluent length and consumed energy of μAPPJ (**section 4.3**), ionization wave dynamics and propagation velocity (**sections 4.4**, **4.5**), and optical emission spectroscopy analysis of μAPPJ (**section 4.6**).

### *4.1 Electrical properties*

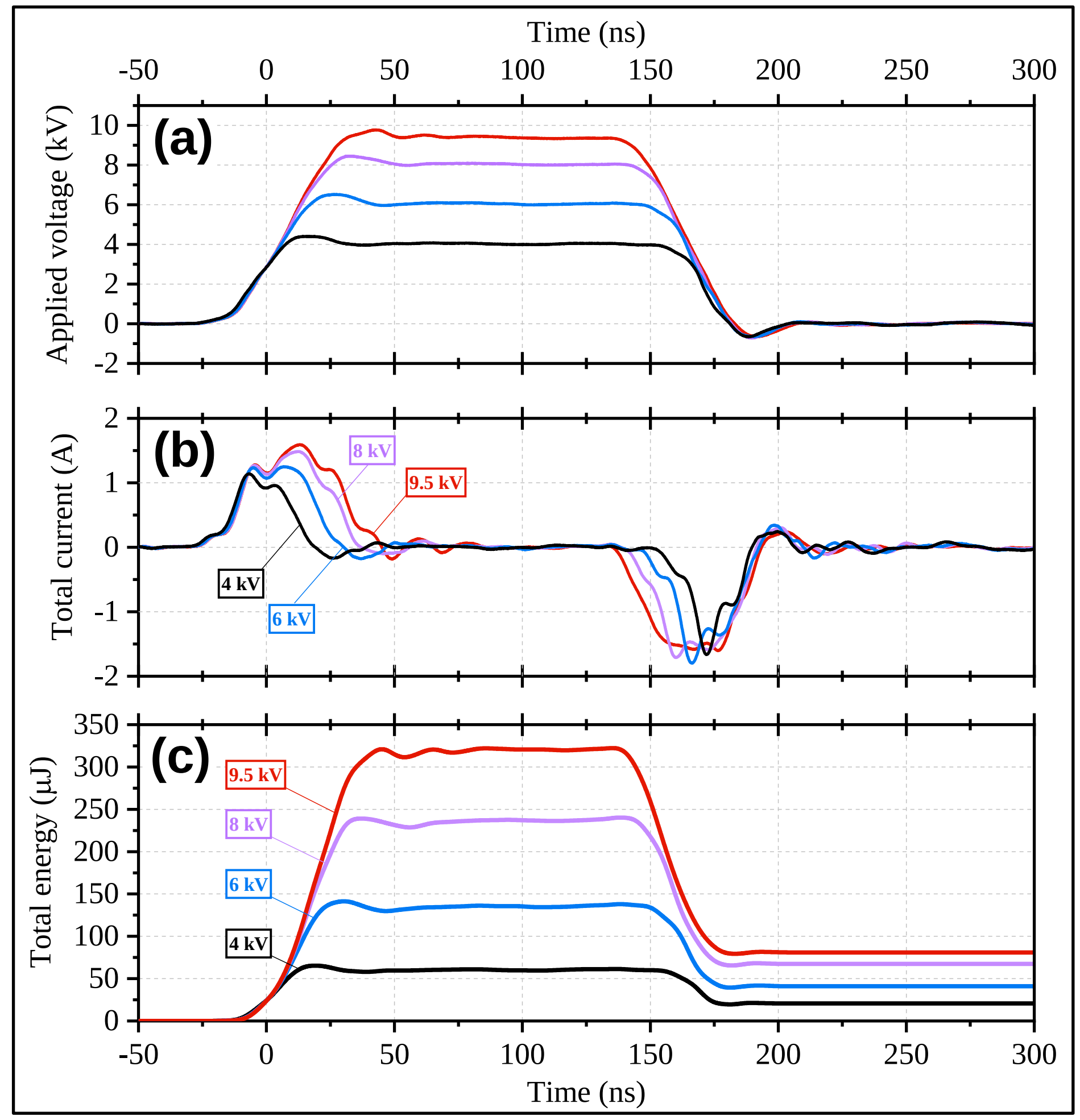

***Figure 4****. Indicative waveforms (100 accumulations in the oscilloscope) of (a) applied high voltage, (b) total current, and (c) total instantaneous energy for different applied voltage amplitudes.*

**Fig. 4(a)** illustrates representative voltage waveforms for the amplitudes ($V_p$) considered in this work: $V_p$=4, 6, 8, and 9.5 kV. In all cases, the repetition frequency ($f_{pulse}$=10 kHz) and the pulse width ($\Delta t_{pulse}$=160 ns) are maintained constant. The corresponding total current signals ($I_T$), recorded while the plasma is "ON" ($Q_V$=0.3 slm), are shown in **Fig. 4(b)**. $I_T$ is the sum of the discharge conduction current ($I_D$) and the displacement/capacitive current ($I_C$) resulting from the inherent capacitive characteristics of the system (μAPPJ, probe and cables). In our conditions, $I_C$ is the dominant current component. This is verified by recording all current waveforms without helium gas in the tube (i.e., plasma "OFF") while varying $V_p$ between 4 and 9.5 kV.

Using the applied voltage and total current waveforms from **Figs. 4(a)** and **4(b)**, the corresponding instantaneous total power is calculated [71]; its integration over a voltage period provides the total energy per pulse (**Fig. 4(c)**). This increases with $V_p$, as expected, which also influences the morphology of the current waveforms. During the rising edge of the voltage, the current is positive; it becomes broader and its peak noticeably increases with higher $V_p$. During the voltage falling edge, the current pulse is negative; it also broadens with increasing applied voltage, though its peak value does not show a noteworthy change. The broadening of the current pulses is a consequence of the intrinsic characteristics of the present power supply, which generates HV waveforms with a nearly constant rise and fall rate (about 200 V/ns). Consequently, larger peak voltages lead to longer rise/fall times, which are reflected on the induced current.

The behavior of the total current is consistent with other studies confirming distinct discharge mechanisms during the rising and falling parts of the voltage waveform [4,14,48,62]. In the voltage rising phase, the positively-biased needle attracts free electrons, which accelerate in the helium channel and enhance the excitation/ionization processes of the gas. In this case, the discharge formation resembles a cathode-directed streamer in the form of a fast ionization wave front propagating from the dielectric tube into the ambient air and depositing charges on the inner capillary surface. In the voltage falling phase, as the voltage amplitude quickly drops to zero, a secondary discharge is formed due to the release of charges that were deposited on the inner surface of the dielectric during the previous discharge, resulting in the observed negative current pulse. The dynamics of each discharge are further captured and confirmed via ICCD imaging (see below).

### *4.2 Averaged discharge morphology and gas flow field*

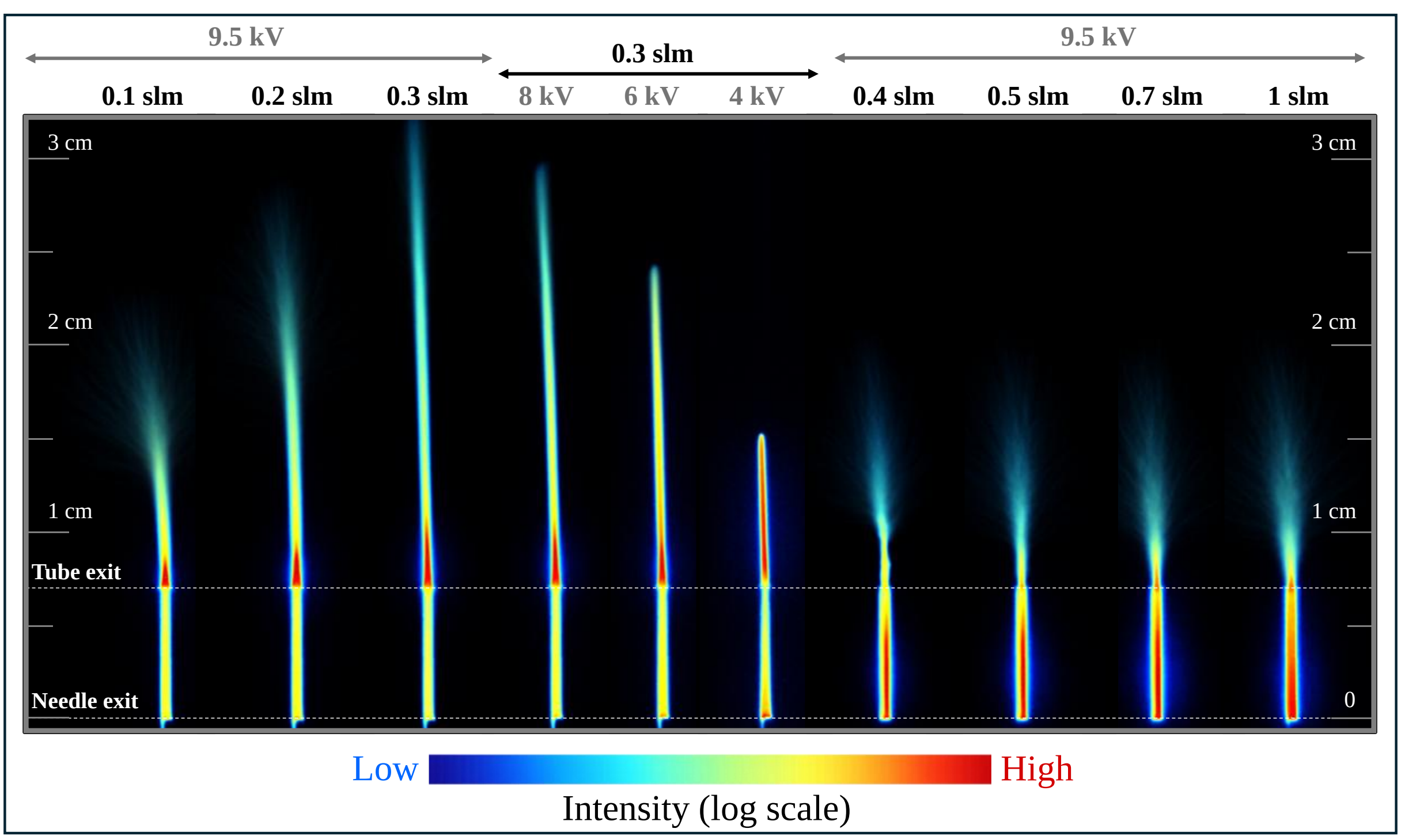

***Figure 5****. ICCD images post-averaged over 500 single-shot frames (ICCD gate width per frame: 300 ns) of the UV-VIS-NIR effluent pattern versus the helium flow rate ($V_p$=9.5 kV, $f_{pulse}$=10 kHz, $\Delta t_{pulse}$=160 ns) and the applied voltage amplitude ($Q_V$=0.3 slm, $f_{pulse}$=10 kHz, $\Delta t_{pulse}$=160 ns).*

**Fig. 5** shows representative false-color ICCD images of the discharge UV-VIS-NIR emission within the dielectric tube (90% transmittance) and after the tube exit, for $Q_V$=0.1–1 slm (at $V_P$=9.5 kV) and $V_P$=4–9.5 kV (at $Q_V$=0.3 slm). These images are obtained by post-averaging 500 single-shot frames with the ICCD gate set to 300 ns. They reveal a significant influence of $Q_V$ on the visible morphology of the effluent, as in other studies [28,36]. At $V_P$=9.5 kV, as the flow rate increases from 0.1 to 0.3 slm, the maximum propagation length increases from approximately 15 mm to 27 mm. At $Q_V$=0.1 slm, the effluent follows a slightly bent path and exhibits a bush-like structure at its tip, indicating that the helium stream weakens rapidly after the exit nozzle and cannot efficiently displace the surrounding air over large distances. This is confirmed by a similar but less pronounced bending and "bush-like" structure observed at $Q_V$=0.2 slm, with a noticeably larger total length compared to 0.1 slm due to the higher flow rate. Upon further increasing the flow rate to 0.3 slm, the jet becomes straighter and more focused, indicating deeper penetration of the helium channel into ambient air, which facilitates ionization wave propagation [36]. At this flow rate, the effluent preserves its morphology at lower voltage amplitudes as well, which primarily affect its length. Indeed, as $V_P$ decreases from 9.5 to 4 kV, the effluent length decreases from approximately 27 to 8 mm, while still remaining straight and collimated across all voltage amplitudes. These results show a stronger effect of the gas flow rate versus applied voltage amplitude on the stability and collimation of the effluent.

For $Q_V \leq$0.3 slm in **Fig. 5**, the luminous effluent is characterized by two distinct axial zones: *(i)* a bright red cone extending from ≈2 mm (after the tube exit) at $Q_V$=0.1 slm to ≈4 mm at $Q_V$=0.3 slm, and *(ii)* a yellow–green tail following the cone, whose length also scales with the flow rate. These results agree with previous studies on He mAPPJs operating at higher flow rates ($Q_V$>2 slm), where the bright cone was linked to He metastable state formation leading to enhanced ionization after the tube exit [85]. In that study, these two emission zones merged below a critical applied voltage amplitude (≈4.5 kV); however, at higher voltage amplitudes, they remained distinct. This is consistent with our experiments, where the cone length progressively increases with decreasing voltage until the two zones merge at $V_P \leq$4 kV.

Furthermore, an abrupt morphological transition occurs when $Q_V$ increases from 0.3 to 0.4 slm. While the cone and the tail regions are still observable but to a much lesser degree, the maximum emission intensity shifts primarily to the interior of the dielectric tube. The effluent becomes wavy and significantly shrunk, while the yellow–green tail takes on a broader "bush-like" form. Notably, at the transitional flow rate of 0.4 slm, the tail occasionally appears bifurcated, an effect observed by the naked eye and captured via DSLR camera (not shown in **Fig. 5**), which is also observed in other studies [33,86]. The visibility of this bifurcation also depends on the angle of observation of the effluent emission by the camera. For $Q_V \geq$0.5 slm, the bifurcation disappears, and the effluent progressively becomes more spread as the flow rate increases up to 1 slm. Overall, these results highlight the critical role of operating parameters (gas flow rate, voltage amplitude) on discharge dynamics. In this study, $Q_V$=0.3 slm and $V_P$=9.5 kV emerged as the optimal operating condition for a stable, collimated, and long effluent in ambient air.

Overall, for $Q_V \geq$0.3 slm a regime seems to be established where shear instabilities appear to govern the effluent behavior. The sudden morphological transition of the effluent at $Q_V \geq$0.4 slm is characteristic of flow destabilization, likely driven by increased jet momentum and/or enhanced entrainment of ambient air [28,36]. Stronger interaction between the helium core and surrounding air promotes shear instabilities at the jet boundary, eventually leading to branching, broadening, and reduced axial consistency.

The ICCD images in **Fig. 5** represent the post-averaged effluent emission from 500 HV pulses. These qualitatively reveal that the helium flow rate strongly influences the global discharge path in ambient air. To better visualize the gas flow field outside the tube, schlieren imaging is performed. **Fig.**

**6** shows the schlieren results for four representative gas flow rates (averaged images over 200 HV pulses). These are captured under two distinct conditions: *(i)* the flow of pure helium in the absence of plasma (plasma "OFF"), and *(ii)* the discharge ignited and propagating inside the dielectric tube and into the ambient air (plasma "ON"). Contrary to **Fig. 5**, schlieren imaging reveals how the helium jet structure, shear-layer, and overall flow stability evolve with increasing flow rate, while also demonstrating how the discharge modifies the structure of the helium channel (see, e.g., the frames corresponding to $Q_V$=0.5 slm).

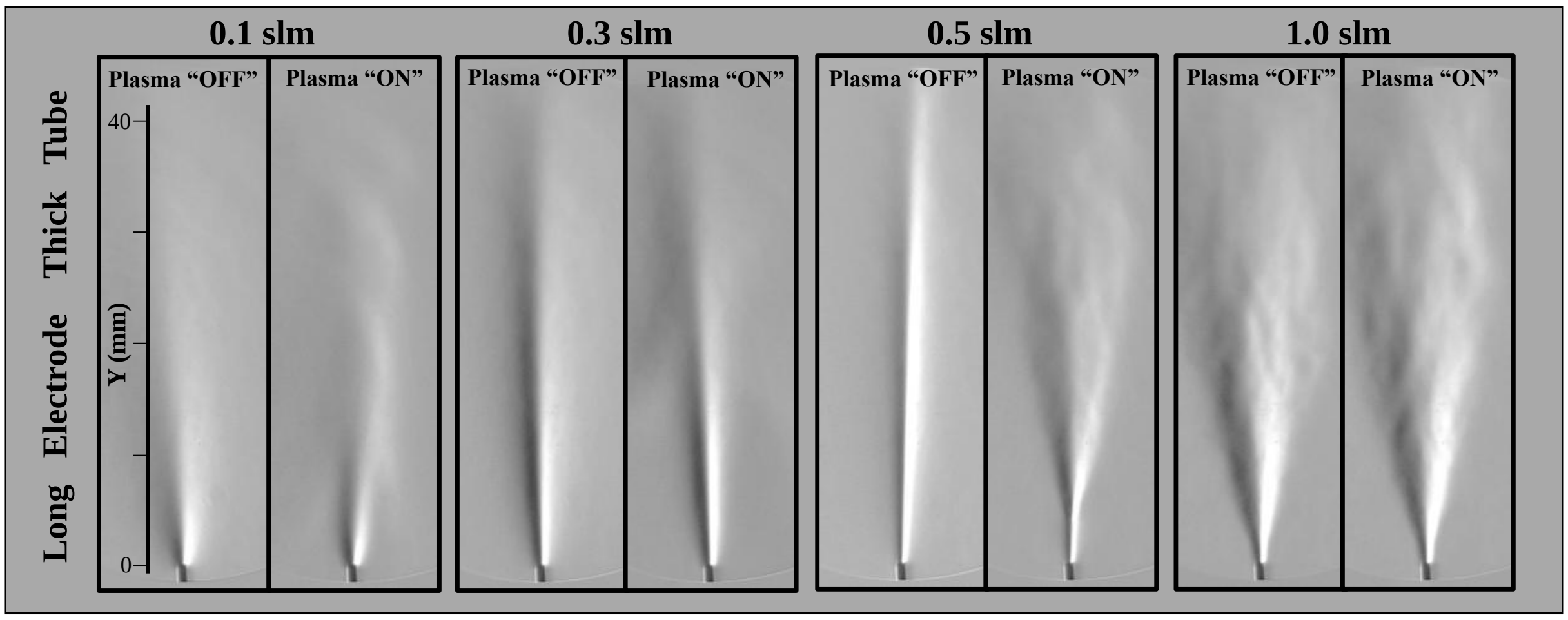

***Figure 6****. Schlieren images captured with a DSLR camera (20 ms exposure time=200 HV pulses), showing the effect of the helium flow rate on the gas dynamics during plasma "OFF" (left frames) and plasma "ON"(right frames) conditions ($V_p$=9.5 kV, $f_{pulse}$=10 kHz, $\Delta t_{pulse}$=160 ns).*

The flow patterns in **Fig. 6** reveal differences in the width of the jet core, the structure of the shear layer, the onset of flow instabilities, and the intensity of turbulence. Similar results have been reported in the literature for different mAPPJs [30,32–34,45,86]. At the lowest flow rate ($Q_V$=0.1 slm), when the plasma is "OFF", the helium jet appears weak and rapidly spreads laterally after exiting the nozzle, reflecting weak flow momentum transfer and strong entrainment of ambient air. The shear layer is thick and diffuse, indicating fast mixing and a short persistence of a pure helium core [36]. When the plasma is "ON" the core of the jet becomes slightly more defined, but the flow momentum seems to remain weak because the jet does not extend as far as for $Q_V$=0.3 slm. Small vortical structures appear farther downstream, which could lead to the "bush-like" morphology observed at the tip of the effluent in the average emission pattern (**Fig. 5**). Buoyancy effects can also become relevant in this regime, as the lighter helium interacts with heavier ambient air, contributing to upward curvature or lateral spreading depending on local mixing conditions.

At $Q_V$=0.3 slm, when the plasma is "OFF", the helium jet becomes remarkably straighter and more collimated. The mixing layer appears thinner and extends farther downstream, indicating improved He flow momentum balance against ambient air entrainment. When the plasma is "ON", the effluent shape is mostly preserved, though small disturbances appear toward its tip, forming faint vortices and slightly reducing its effective length. In this regime, the length reaches a maximum of about 25 mm, consistent with **Fig. 5**.

At $Q_V$=0.5 slm, when the plasma is "OFF", the jet shows an even longer and narrower core compared to 0.3 slm, reflecting increased flow momentum. However, when the plasma is "ON", the flow becomes noticeably unstable. This makes clear that the ignition of plasma significantly perturbs the mixing layer and enhances shear in this flow rate, causing early onset of instabilities just a few millimeters downstream of the nozzle. Only a short and collimated zone of a few millimeters remains;

beyond this point, the jet rapidly expands into a turbulent, conical plume. This behavior correlates directly with the pronounced expansion and shrinking observed in the corresponding emission pattern in **Fig. 5**, consistent with other works referring to mAPPJs [30].

Finally, at $Q_V$=1 slm, it is remarkable that even when the plasma is "OFF", the helium jet is already turbulent, exhibiting large fluctuations and rapid conical spreading immediately after the nozzle. These cause a characteristic "hissing noise" that is easily perceived by ear when He gas flows from the dielectric tube into the surrounding air. Unlike at lower flow rates, at $Q_V$=1 slm there is no clearly defined collimated zone. When the plasma is "ON", only a slight qualitative change is observed in the large-scale morphology. The gas flow is already dominated by hydrodynamic instabilities, and the plasma-induced perturbations seem negligible compared to those inherent in the high-momentum unstable jet.

### *4.3 Single-shot discharge morphology, luminous effluent length and consumed electric energy*

**Figs. 5** and **6** provide time-averaged views of the effluent emission profile, but they do not elucidate the underlying fast discharge dynamics occurring in much smaller timescales (<300 ns). To address this, we first performed ICCD imaging in single-shot mode, where the total emission over a single voltage pulse (gate width: 300 ns) is captured. Representative single-shot images of the effluent are shown in **Fig. 7** for different gas flow rates and applied voltage amplitudes.

A comparison between the single-shot images at $Q_V$=0.1 (**Fig. 7(a1)**) and 0.3 slm (**Fig. 7(a2)**) reveals stochastic lateral branches which differ in number, inception point and spatial distribution. At the lowest flow rate (0.1 slm), the discharge core is significantly shorter and exhibits an early onset of stochastic branching starting as close as 5 mm from the nozzle. These branches appear widely divergent and lack a dominant direction, resulting in the "bush-like" morphology previously identified in the time-averaged profiles (**Fig. 5**). This behavior, captured for the first time in He µAPPJs, is attributed to the low momentum of the helium stream, which allows ambient air to easily penetrate its core (**Fig. 6**), creating a highly irregular and unstable helium–air interface that the IWs are forced to follow.

In contrast, the single-shot images corresponding to $Q_V$=0.3 slm uncover a nearly vertical effluent with improved collimation. The central core remains stable and unbranched for at least 15–20 mm downstream of the exit nozzle. Lateral stochastic branches still exist in this regime, but they are less frequent and occur much farther downstream compared to the 0.1 slm. This confirms that at 0.3 slm, the ionization wave propagation is better shielded from the perturbation of ambient air. Consequently, the contribution from stochastic lateral branches on the morphology of the averaged luminous effluent pattern (**Fig. 5**) is less significant than for $Q_V$<0.3 slm. This feature correlates with the corresponding schlieren results recorded on a similar time scale (**Fig. 6**), where the helium gas jet is straighter, more focused and minimally perturbed at $Q_V$=0.3 slm. Thus, at $Q_V$=0.3 slm, a coherent, helium-rich core is maintained enabling a sufficiently long and repeatable luminous effluent with minimal branching.

At $Q_V$=0.5 slm (**Fig. 7(a3)**), the single-shot images show a much higher degree of instability than for $Q_V \leq$0.3 slm. A very short (<3 mm) distorted/wavy central core is barely visible, which rapidly splits into multiple stochastic lateral branches. These branches vary significantly in length, deviation angle, and number from shot to shot. The resulting global pattern is wider and more irregular than at 0.3 slm. Using single-shot ICCD images with a 150 ns gate width (**Figs. 7(a3.i)** and **7(a3.ii)**), we confirmed that these branches develop exclusively during the rising voltage edge (**Fig. 7(a3.i)**); in contrast, the falling edge (**Fig. 7(a3.ii)**) is characterised by a diffuse effluent extending only for a few millimeters after the exit nozzle. This indicates distinct discharge formation mechanisms during these two voltage phases, as already discussed in **Fig. 4(b)**. The average profile at $Q_V$=0.5 slm shows a diffuse, broadened emission region, along with a wavy core and an absence of a long dominant trajectory (**Fig. 5**). This closely matches the corresponding schlieren image in **Fig. 6**, where turbulence and shear-layer distortions emerge just a few millimeters downstream of the exit nozzle. The coincidence of branching onset (**Fig. 7**) with the turbulent transition region (**Fig. 6**) suggests that the discharge follows locally favorable channels generated within unstable helium-air mixtures.

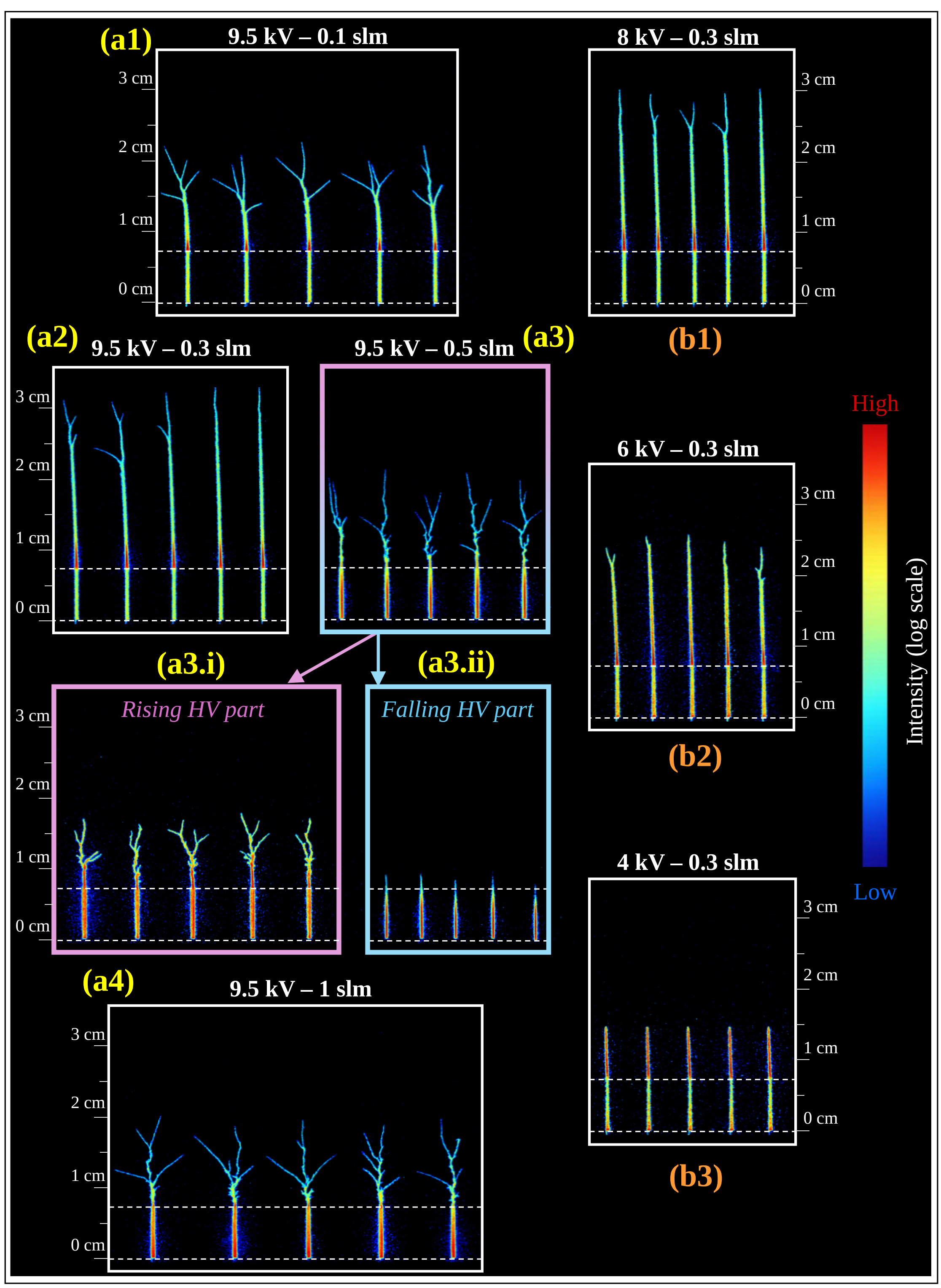

***Figure 7***. *Representative single-shot false-color ICCD images (gate width: 300 ns) of the UV-VIS-NIR effluent pattern versus the He flow rate (**a1–a4**; $V_P$=9.5 kV) and the voltage amplitude (**b1–b4**; $Q_V$=0.3 slm). Frames **a3.i** and **a3.ii** depict single-shot images taken during the rising (150 ns gate width) and falling (150 ns gate width) phases of the voltage pulse ($V_P$=9.5 kV and $Q_V$=0.5 slm).*

Finally, at $Q_V$=1 slm (**Fig. 7(a4)**), the single-shot ICCD images show even stronger branching activity. The central core of the effluent becomes very short (if not absent at all) and surrounded by a broad random distribution of ionization paths. In this case, short branches can even appear just after the tube exit, indicating increased instability of the gas channel. The corresponding averaged image in **Fig. 5** is diffuse and dispersed, reflecting the absence of any reproducible and long ionization trajectory. This is again consistent with the schlieren results (**Fig. 6**), showing a highly distorted helium jet with rapid air entrainment and no stable, collimated and extended propagation path. At this flow rate, the helium-air mixture is so strongly disturbed that no coherent path remains for persistent ionization wave propagation.

**Figs. 5**, **6** and **7** collectively indicate $Q_V$=0.3 slm as a critical operating flow rate threshold, that provides the most stable helium core and the longest and most reproducible propagation path (when plasma is "ON"). Below this value, flow momentum is insufficient, causing the jet to bend and spread; above it, hydrodynamic- and/or plasma-induced instabilities critically disrupt the helium channel, leading to the fragmentation of the ionization wave path.

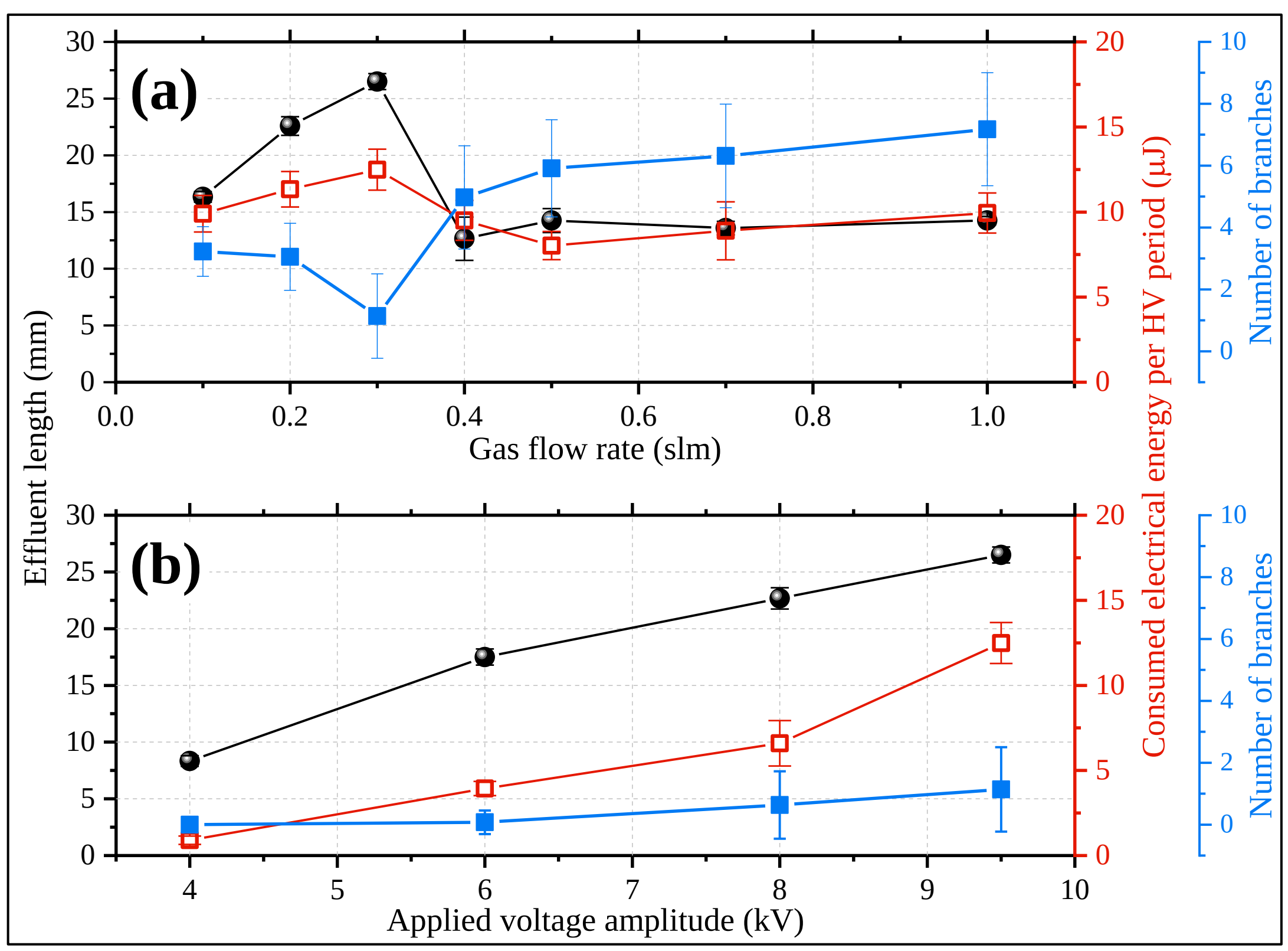

***Figure 8****. Effect of (a) helium gas flow rate ($V_P$=9.5 kV, $f_{pulse}$=10 kHz, $\Delta t_{pulse}$=160 ns) and (b) applied voltage amplitude ($f_{pulse}$=10 kHz, $\Delta t_{pulse}$=160 ns, $Q_v$=0.3 slm), on the effluent length (black), consumed discharge energy per HV period (red), and number of detected branches (blue). Error bars on branch number represent the standard deviation calculated from 500 single-shot ICCD images recorded under each condition. Error bars on energy are obtained from 5 series of experiments performed during two different days (averaged over 100 signals). Length error bars are obtained from 500 single shot ICCD images.*

The weaker branching activity observed at $Q_V$=0.3 slm is further analyzed as a function of $V_P$ (**Figs. 7(a1)** versus **7(b1) – 7(b3)**). As the voltage amplitude decreases from 9.5 (**Fig. 7(a1)**) to 8 (**Fig. 7(b1)**) and then 6 kV (**Fig. 7(b2)**), the number and length of the branches are progressively reduced, until they completely disappear at $V_P$=4 kV (**Fig. 7(b3)**). In this low-voltage regime, the discharge is confined to a single, short and collimated central core. Correspondingly, the total effluent length decreases significantly with voltage, becoming less than 10 mm at $V_P$=4 kV. These observations suggest that higher voltage amplitudes generate stronger ionization wave fronts that can amplify small fluctuations in the helium-air boundary layer. This increases the sensitivity of the effluent to instabilities, thereby promoting stochastic branching at the tip of the effluent. In pulsed plasma jets, even a single voltage pulse can trigger a turbulent front that convects downstream at the gas velocity [34,44,45].

**Fig. 8** shows the influence of both the gas flow rate and the applied voltage amplitude on the luminous effluent length, the consumed discharge energy per HV period, and the number of lateral branches. The effluent length is obtained from spatially-calibrated ICCD images such as in **Figs. 5** and **7** (using a scale of 1 mm per 30 px determined by taking photos of a ruler placed at the same position as the capillary tube), while the number of branches is obtained using a dedicated Matlab® routine. This allows user-guided branch selection/detection followed by automated counting and data storage. It should be noted that the measurement of the number of branches here refers to the 2D field of view of the ICCD camera. This being said, we could not perform a 3D analysis in this work which could provide a more accurate picture of the 3D branch morphology. The consumed discharge energy per pulse is calculated as the net difference between the energies measured under plasma "ON" and plasma "OFF" conditions (as described in **Fig. 4**) for each flow rate and HV amplitude. In this case, the voltage and current waveforms used are averaged over 500 signals. To ensure that no discharge is present under the plasma "OFF" cases, we took ICCD images at different time instants during the HV pulse, without observing any light emission.

**Fig. 8(a)** quantitatively illustrates the critical role of the helium flow rate on the luminous effluent length and stability. By increasing $Q_V$ between 0.1 and 0.2 slm, the length (black circles) increases from 16 to 22 mm, respectively, and reaches a maximum of approximately 27 mm at $Q_V$=0.3 slm (as also seen in **Figs. 5** and **7**). This behavior is consistent with a rise in the corresponding energy consumed by the discharge, and a decrease in the average number of branches (blue squares) from 3.2 ($Q_V$=0.1 slm) to 3.1 ($Q_V$=0.2 slm) to 1.1 ($Q_V$=0.3 slm). The smallest branch number at 0.3 slm confirms that a stable, well-collimated helium core suppresses stochastic behavior. At $Q_V$=0.4 slm, an abrupt transition occurs in the effluent length which drops sharply to about 12.5 mm, while the number of stochastic branches increases noticeably (≈5 branches per voltage pulse on average). This transition marks the onset of important instabilities and ambient air entrainment for $Q_V \geq 0.4$ slm, which disrupts the ionization channel. Interestingly, while the consumed discharge energy per HV period (red squares) peaks at 0.3 slm (≈12.5 μJ on average), it falls below 10 μJ at higher flow rates. The disruption of the helium channel above 0.3 slm limits the total volume and duration of the discharge, thereby reducing the total energy coupled to the luminous effluent despite the high applied voltage. The length and consumed energy remain quasi-stable in the range $0.4 \leq Q_V \leq 1$ slm. However, the average number of branches increases slightly, most probably due to enhanced hydrodynamic effects and mixing with ambient air (**Fig. 6**).

**Fig. 8(b)** demonstrates that for a fixed optimal flow rate ($Q_V$=0.3 slm), all parameters scale positively with the applied voltage amplitude. Both the luminous effluent length and the consumed discharge energy exhibit a steady, near-linear increase as $V_P$ rises from 4 to 9.5 kV. This is expected as the effluent remains well-collimated in this voltage range (**Fig. 5**), while higher applied voltages induce a stronger electric field to sustain a longer propagation of the ionization wave into the ambient air [14,18,28]. Simultaneously, the average number of branches slightly increases with voltage amplitude as also qualitatively observed in **Fig. 7**. This supports the hypothesis that higher voltage amplitudes generate more intense ionization wave fronts that are more susceptible to amplifying infinitesimal fluctuations at the helium–air interface, leading to the formation of stochastic ionization paths (branching) at the tip of the effluent even within a relatively stable gas flow regime.

### 4.4 *Space- and time-resolved ionization wave propagation*

The variations in the gas flow rate and the voltage amplitude affect the velocity of IWs and their capacity to penetrate deeper into the ambient air. To investigate these fast dynamics, space- and time-resolved ICCD imaging is performed allowing for the accurate tracking of the ionization wave front evolution as a function of both parameters.

**Fig. 9** presents the spatio-temporal development of the discharge emission for four representative flow rates, captured with a 2 ns time resolution. These images track the propagation of the total light of the ionization wave front as it travels from the needle tip (z=0), through the glass microtube (0<z≤7 mm), and into the ambient air (z>7 mm). This analysis is possible because the discharge inception jitter is extremely low (<1 ns).

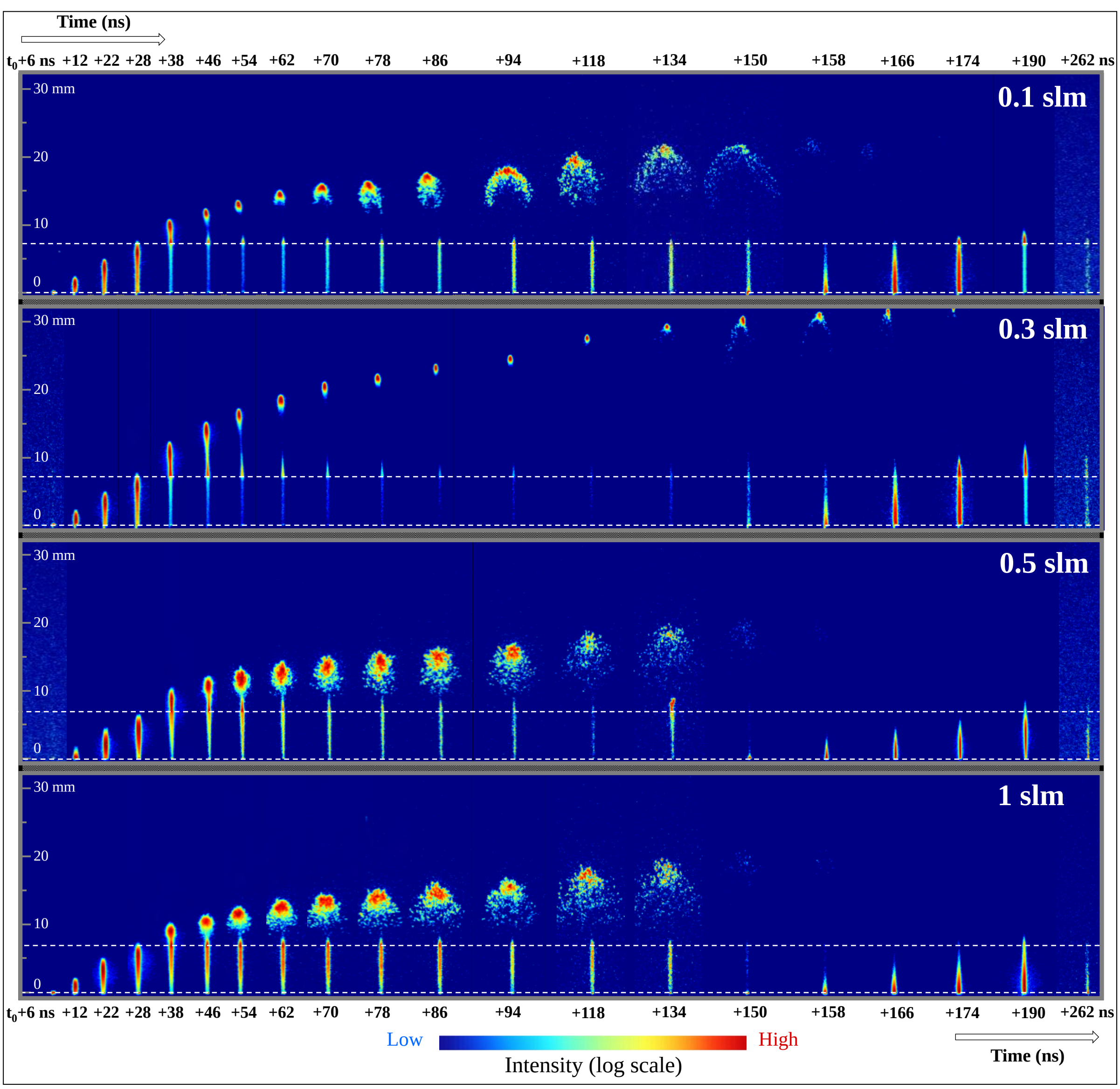

***Figure 9**. ICCD images (2 ns gate width; 100 accumulations; 2 ns step; gain: 230) at $V_P$=9.5 kV capturing the ionization wave front propagation inside the microtube and in the ambient air for different helium gas flow rates. The instant $t_0$=0 ns is when the ionization wave front first appears at the tip of the needle (indicated by the bottom horizontal dashed line; the top horizontal dashed line indicates the exit nozzle of the dielectric tube).*

Between time instants $t_0$=0 and $t_0$+28 ns, the IW propagates inside the dielectric tube; its dynamics appears almost identical for all flow rates, characterized by a well-defined, intense ionization wave front. However, upon exiting the nozzle (upper dashed horizontal line in **Fig. 9**), the IW dynamics diverge significantly depending on the gas flow rate. At 0.1 slm, the wave front propagates into the air but begins to broaden and decelerate relatively early, reaching a maximum height of approximately 15 mm. The ionization wave front becomes increasingly spread, mirroring the weak gas momentum and rapid air entrainment indicated by the schlieren images (**Fig. 6**). At 0.3 slm, the IW exhibits the most stable and extended propagation. The ionization wave front remains highly collimated and intense, reaching the maximum distance (around 27 mm). The sustained velocity and focused morphology confirm that this flow rate provides a long, laminar helium channel that effectively shields the IW from the surrounding air perturbation. At 0.5 slm, a drastic change is observed. Almost immediately after the nozzle exit, the IW front undergoes rapid lateral expansion, progressively becoming a broad cloud. The maximum propagation height (10–15 mm) is severely limited again compared to $Q_V$=0.3 slm. This correlates with the onset of shear-layer instabilities and turbulence noted in **Fig. 6**, induced by the plasma ignition; the rapid mixing with ambient air increases the local instabilities and breaks the coherent propagation channel, forcing the IW to terminate prematurely. This is also happening at 1 slm, which exhibits a relatively broader cloud-shaped emission profile.

At 0.3 slm the IW front in air appears to be detached ($t>t_0$+54 ns) from a residual emission channel behind it. After its detachment, the backward emission inside the tube decreases rapidly and seems to extinguish roughly after $t_0$+86 ns, while the IW front still propagates. While a detachment is also seen at 0.1 slm and, to a lesser degree, at $Q_V$>0.3 slm, a rather intense backward emission in the tube is still observable for these flow rates until roughly at $t_0$+150 ns, i.e., before the occurrence of the falling edge of the voltage pulse. This could mean that at 0.3 slm a significant portion of the consumed energy (**Fig. 8**) is delivered to the ionization wave front, which favors its propagation over much longer distances.

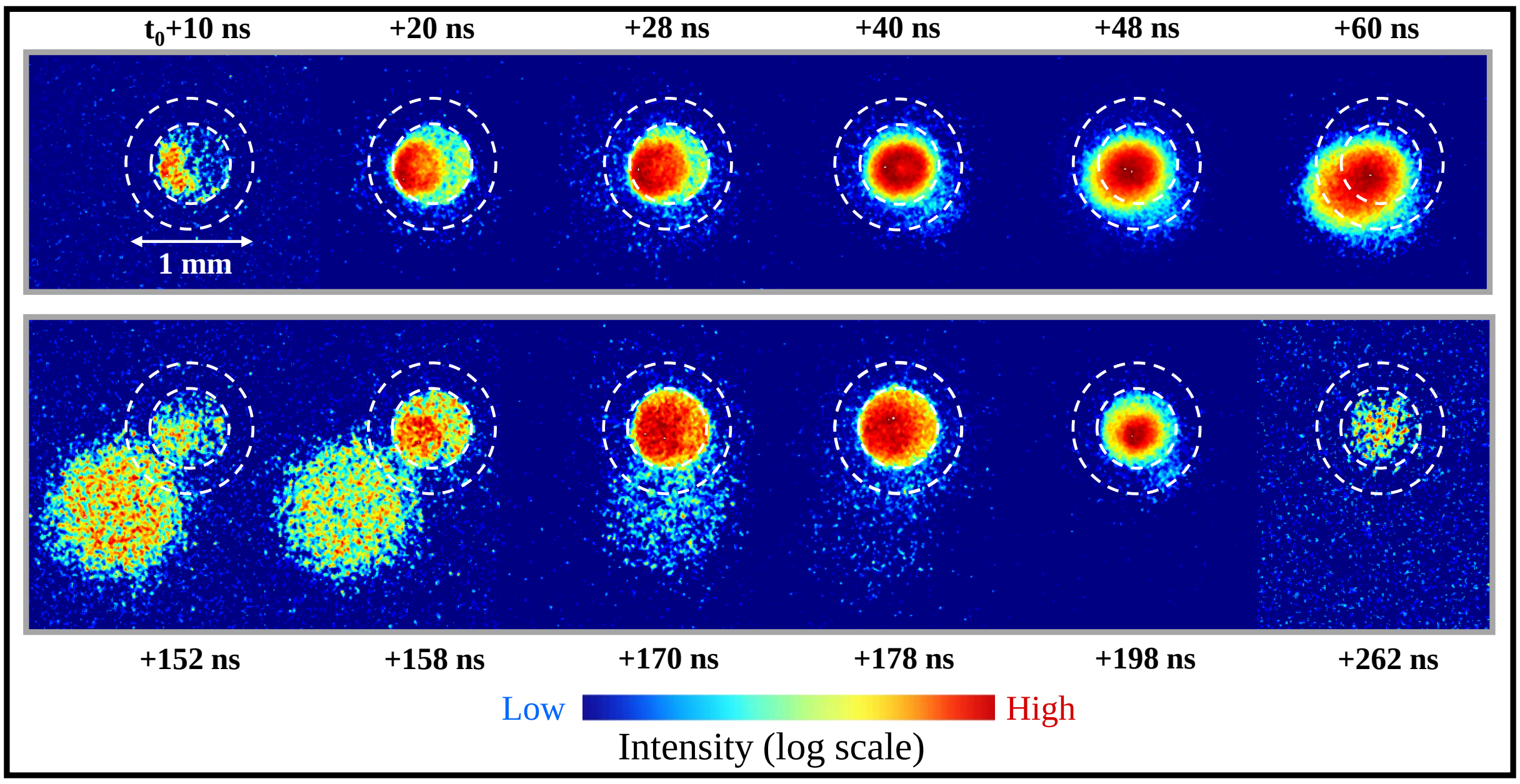

***Figure 10***. *ICCD images (2 ns gate width; 100 accumulations; 2 ns step; gain: 230) capturing the ionization wave front propagation inside the microtube and in the ambient air at $V_P$=9.5 kV and $Q_V$=0.3 slm.*

Finally, the frames corresponding to the latest time instants ($t>t_0$+150 ns) reveal a secondary, diffuse emission which corresponds to the falling edge of the voltage pulse. This originates at the tip of the needle inside the tube, extends downstream, and terminates only a few millimeters above the dielectric tube exit. At $Q_V$=0.3 slm, this emission co-exists with the primary ionization wave front until at least $t_0$+174 ns (primary IW moves out of the field of view of the ICCD). Contrary, the primary ionization wave fronts corresponding to the other flow rates have already vanished by this time instant. During this phase, the release of surface charges previously deposited on the dielectric walls triggers a secondary discharge (similar to DBDs). Unlike the primary IW associated with the rising voltage edge, this secondary phase does not exhibit a high-velocity IW front (**section 4.5**); instead, it manifests as a slower spatially expanding, decaying glow. This underlines the distinct physical mechanisms governing the two phases of the pulsed discharge discussed on **Fig. 4**.

**Fig. 10** provides a complementary end-on (transverse) view of the ionization wave dynamics at the optimal flow rate, $Q_V$=0.3 slm. This perspective allows for a more complete view of the transverse discharge structure during both the rising and the falling edges of the voltage pulse, being directly correlated with the side-on observations in **Fig. 9**. The transition from an annular structure inside the tube to a solid-core IW front in the effluent is seen which highlights the importance of the gas-dielectric interface for IW initiation and subsequent stabilization of the discharge channel.

During the initial propagation instants inside the microtube ($t<t_0$+20 ns), the emission exhibits an annular (hollow) structure, revealing that the ionization wave front is first concentrated near the dielectric wall. This could be due to local electric field enhancement at the gas-dielectric interface, potentially exacerbated by a slight off-centering of the needle within the dielectric tube. As the IW front reaches the nozzle exit at $t_0$+28 ns, the emission fills the central volume, evolving into a solid-core IW front. Once in the ambient air (from $t_0$+40 ns to $t_0$+60 ns), the IW front maintains a solid core, even as it progressively follows a slightly off-axis propagation path. Notably, at approximately $t_0$+60/62 ns for both 0.1 and 0.3 slm, the ionization wave front becomes detached from the main discharge channel (**Fig. 9**), continuing its propagation as a discrete structure. This detachment, being less evident for $Q_V$>0.3 slm, reflects the transition from a continuous discharge column behind the IW front to a self-propagating wave front. The persistence of this solid-core, detached front at 0.3 slm confirms that the IW effectively utilizes the entire helium-rich channel in this case, accounting for the high intensity and stable propagation observed in the side-on profiles (**Fig. 9**).

Furthermore, in **Fig. 10** the images at $t_0$+152 ns (first appearance of the secondary discharge at the needle tip) and $t_0$+178 ns (beginning of extinction of the primary discharge) provide additional evidence of the co-existence phase. At $t_0$+158 ns, while the primary IW front (visible as the larger, diffuse cloud) propagates deep into the ambient air far from the nozzle, the secondary discharge intensifies at the center of the tube. The end-on view spatially separates these two distinct physical phenomena occurring at different axial positions. At $t_0$+170 ns, the secondary discharge exits the tube nozzle. Unlike the primary IW, this appears more constricted and remains strictly confined to the central axis (from $t_0$+178 ns to $t_0$+198 ns in **Fig. 10**), while it does not develop into a discrete intense IW front characteristic of the rising edge as shown in **Fig. 9**. Finally, by $t_0$+262 ns, the secondary emission transitions into a low-intensity glow that slowly decays as the voltage returns to zero.

### *4.5 Time-resolved total discharge emission and ionization wave velocities*

For a better correlation of the discharge total emission properties with the electrical signals, **Fig. 11** provides a summary of the discharge temporal dynamics at $Q_V$=0.3 slm, correlating the applied voltage and discharge current signals with the temporal evolution of the spatially-integrated total emission intensity and ionization wave velocity.

The primary IW front is initiated in the tube a few ns after $t_0$=0 ns (**Fig. 11(b)**; black spheres), i.e., when the discharge current starts to rise (**Fig. 11(b)**; blue signal). The current rises fast between $t_0$ and $t_2$=$t_0$+20 ns, inducing a sharp increase in the emission intensity near the needle tip (**Fig. 11(a)**; region (1), black line). As the IW front moves toward the tube exit ($t$<$t_3$=$t_0$+28 ns), it undergoes significant acceleration until $t_2$ where its velocity reaches a short temporary plateau (3.85×10$^2$ km/s). This plateau coincides with a small shoulder observed in the discharge current at $t_2$ which is maintained for about 5 ns (i.e., between $t_2$ and $t_3$). Then, the IW front velocity increases again until a few ns after $t_3$, where it reaches its maximum value overall (about 6×10$^2$ km/s) just after the nozzle exit. The current (**Fig. 11(b)**; blue signal) and total intensity (**Fig. 11(a)**; region (2), red curve) signals also peak around the same instant. These maxima are likely driven by enhanced excitation and ionization processes occurring within the helium–air mixing layer. Subsequently, these three signals decay rapidly until the instant $t_4$ where the maximum overall emission intensity in ambient air (**Fig. 11(a)**; region (3), green curve) is reached. Between $t_4$=$t_0$+40 ns and $t_5$=$t_0$+46 ns, the current starts to rise again leading to a slower deceleration rate of the velocity until the instant $t_6$=$t_0$+60/62 ns where the IW front is detached from the continuous luminous channel behind it (**Fig. 9**). After $t_6$, all signals decay progressively while the IW front propagates as a discrete, self-sustained "plasma bullet" until its extinction.

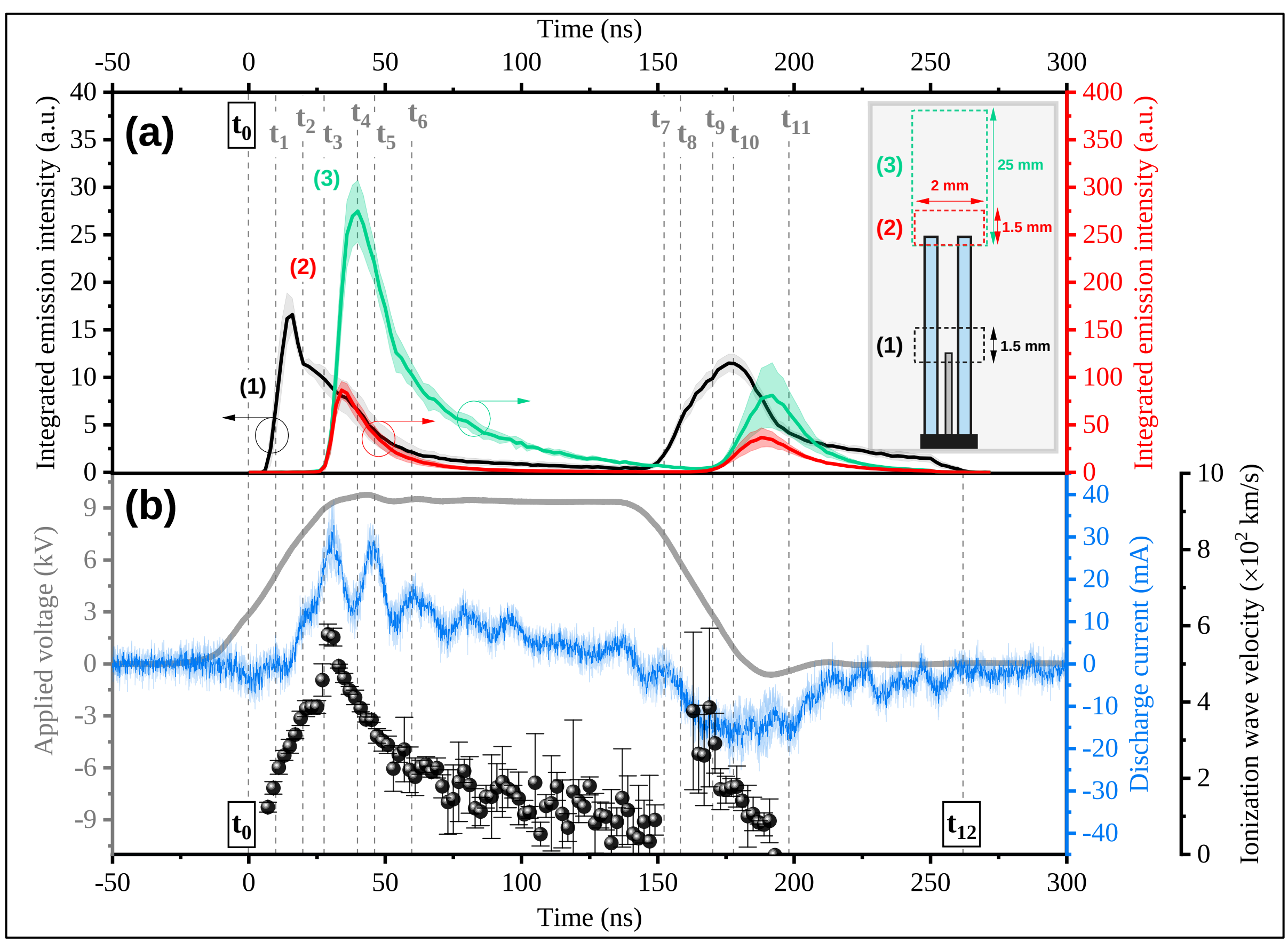

***Figure 11***. ***(a)***: *time-resolved and space-integrated total emission intensity (three series of experiments during different days/months) from three discharge regions shown in the inset: (1) around the needle tip (black dashed rectangle), (2) around the dielectric tube nozzle (red rectangle), and (3) in the ambient air (green rectangle).* ***(b)***: *applied voltage (grey), discharge current (blue), and ionization wave front velocity (black spheres; 5 series of experiments on different dates). Instants $t_0$–$t_{12}$ shown in the figure are commented in the main text.*

During the falling edge of the voltage pulse, the secondary discharge is initiated at the needle tip ($t_7$=$t_0$+152 ns), being triggered by the drop of the HV amplitude and the release of deposited surface charges. This secondary discharge propagates downstream, reaching the nozzle exit at $t_9$=$t_0$+170 ns.

Quantitatively, the secondary discharge is weaker and relatively slower than the primary discharge, while its integrated emission intensity is also lower. The emission remains largely confined to the interior of the tube and the vicinity of the nozzle, confirming the short diffuse morphology observed in the time-resolved images of **Figs. 9** and **10**. In this case, the calculation of the velocity is not as simple and straightforward as for the primary IW front, because the discharge exhibits a wider diffuse head, which leads to larger uncertainties. Nevertheless, for the sake of relative comparison, an indicative calculation is attempted, and the result is shown in **Fig. 11(b)**.

Furthermore, the discharge current depicted in **Fig. 11(b)**, calculated by subtracting the displacement current ($I_C$; Plasma "OFF") from the total measured current ($I_T$; Plasma "ON"; see **Fig. 4**), exhibits two main pulses that are synchronized with the optical emission peaks. The first, more intense positive current pulse corresponds to the formation and propagation of the primary IW, while the second, weaker negative pulse aligns with the secondary discharge. The synchronization between these multi-modal measurements confirms the precision of the experimental timing and reinforces the physical model of a dual-phase discharge governed by distinct inception and propagation mechanisms.

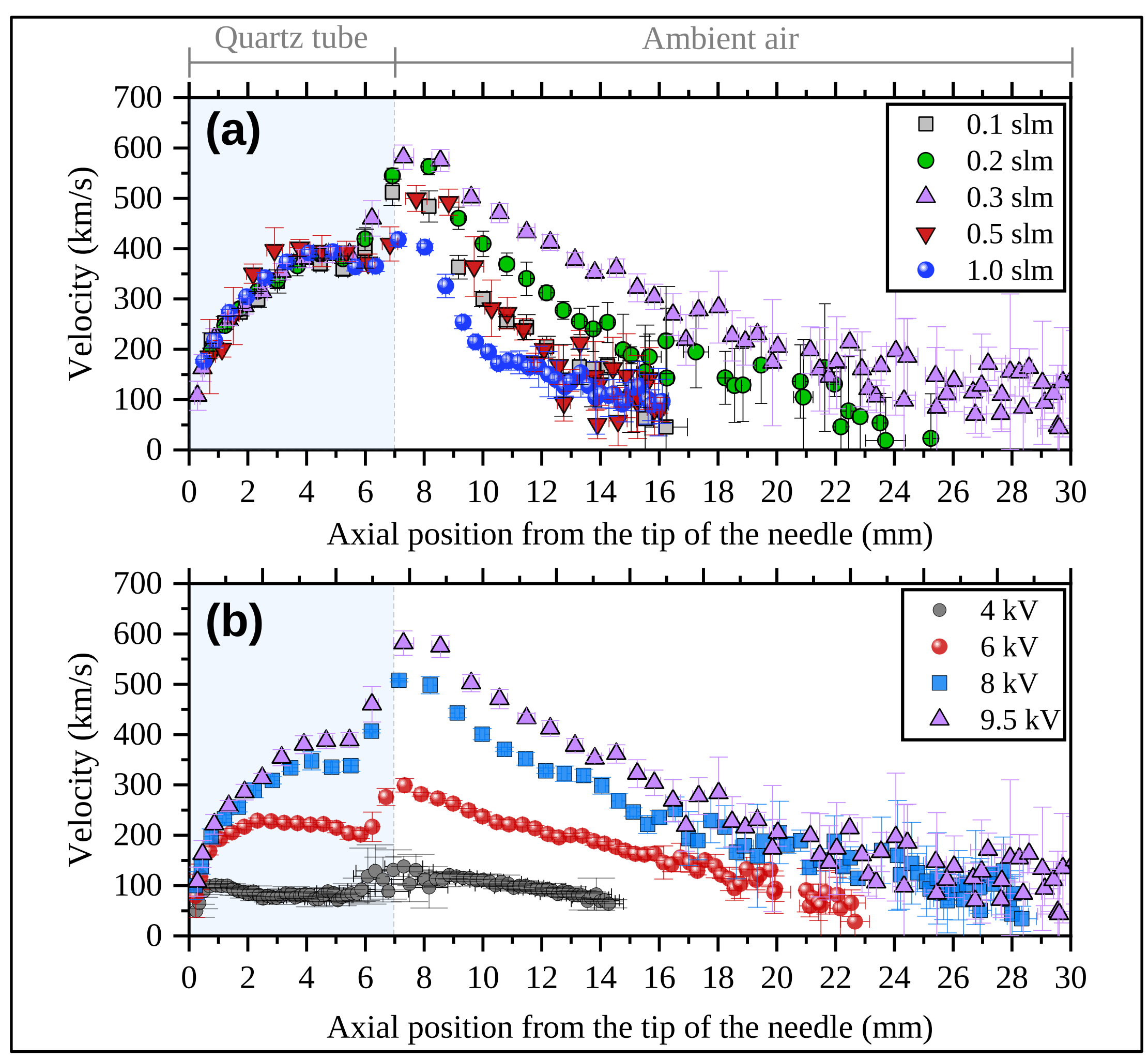

***Figure 12**. Primary ionization wave front velocities along the discharge propagation axis inside the microtube and in the ambient air for different (a) helium gas flow rates ($V_P$=9.5 kV), and (b) applied voltage amplitudes ($Q_V$=0.3.slm). Error bars refer to 5 independent experiments on different days/months.*

**Fig. 12** quantitatively summarizes the primary IW front velocity for various flow rates and voltage amplitudes as a function of the axial position (z), where z=0 corresponds to the needle tip and z=7 mm marks the transition from the microtube to the ambient air (nozzle exit). **Fig. 12(a)** illustrates the effect of the gas flow rate on the IW front velocity at $V_P$=9.5 kV. Regarding the propagation inside the tube, for nearly all operating conditions, the ionization wave front exhibits a similar profile, undergoing a gradual acceleration as it leaves the needle tip and travels toward the nozzle. This acceleration is driven by the electric field enhancement at the needle tip as the voltage increases (**Fig. 11(b)**). Interestingly, the velocity does not increase monotonously; instead, it reaches a distinct plateau between z≈4–6 mm before exiting the nozzle, as discussed in **Fig. 11(b)**. This plateau suggests a regime of local equilibrium, where the rate of energy gained from the axial electric field is balanced by energy losses in the propagation channel. The velocity increases again for distances z>6 mm and reaches its peak value slightly after the nozzle exit. This secondary acceleration is attributed to the increased interaction of the IW front with ambient air species, leading to enhanced excitations and ionizations, most likely via collisions with He metastables triggering Penning ionization of nitrogen molecules [85,87]. This peak is more pronounced at lower flow rates ($Q_V$≤0.3 slm), where the IW front velocity can reach up to $6\times10^2$ km/s ($Q_V$=0.3 slm).

Additionally, **Fig. 12(a)** shows that the flow rate dictates the sustainability of the velocity in ambient air (z>7 mm). In agreement with the previous qualitative results (**Figs. 5** and **7–9**), the optimal case is obtained at 0.3 slm, where the IW front maintains the highest velocity over the longest distance, being still measurable up to z≈30 mm in **Fig. 12**. This confirms the presence of a sufficiently stable, long-reaching helium channel that minimizes the quenching of excited species by ambient air. At lower flow rates ($Q_V$<0.3 slm), while the peak velocities at the nozzle remain relatively high, they decelerate more rapidly than at 0.3 slm, as the lower gas momentum allows for faster mixing with the surrounding atmosphere. In the high flow rate regime ($Q_V$>0.3 slm), despite the similar velocities with $Q_V$=0.3 slm inside the tube, the IW undergoes an abrupt deceleration immediately after the nozzle exit. This drop again correlates with the onset of shear-layer instabilities and turbulence observed in **Fig. 6**; the plasma effect on the gas flow and the rapid entrainment of a substantial amount of air species into the effluent dramatically distort the ionization wave front structure.

**Fig. 12(b)** displays the effect of the applied voltage amplitude on the ionization wave front velocity at the optimal flow rate of 0.3 slm. The plateau behavior revealed in **Fig. 12(a)** is still evident for all voltage amplitudes. Notably, the IW front velocity at lower $V_P$ reaches this plateau over shorter axial distances/times, notably due to the different rise times of the corresponding voltage pulses (**Fig. 4**). At 4 kV, the velocity stabilizes almost immediately (within z<1 mm), whereas at 9.5 kV, the IW front undergoes sustained acceleration before leveling off at around z=4 mm. This correlation also suggests that at lower $V_P$, the lower induced electric field limits the maximum velocity the wave can achieve. For propagation outside the tube, as $V_P$ varies from 4 to 9.5 kV, the corresponding peak velocity at the nozzle exit increases from approximately $10^2$ km/s to about $6\times10^2$ km/s. At the lowest $V_P$=4 kV, the velocity remains nearly constant and relatively small (~$10^2$ km/s) throughout its short trajectory. Stronger externally-applied electric fields at higher voltage amplitudes help maintain the local field strength above the ionization threshold over greater distances that results in the faster ionization wave fronts, as observed in **Fig. 12(b)**, which can penetrate further into the surrounding atmosphere.

### *4.6 Space-resolved and time-integrated optical emission spectroscopy*

To gain further insights into the axial distribution of reactive excited species formed inside the capillary tube and along the effluent, time-integrated OES is performed at the optimal operating condition ($V_P$=9.5 kV, $Q_V$=0.3 slm). It is recalled that the capillary tube has 90% transmittance in the range 250–1000 nm. This is considered small enough to not affect a reliable comparison of the light intensities inside and outside the tube since these differ significantly between them. **Fig. 13** displays the self-normalized axial emission intensity profiles for the representative dominant radiative species detected: (OH(A–X) at

306.4 nm, O I at 777.5 nm, $N_2$(SPS) at 337.1 nm, $N_2^+$(FNS) at 391.4 nm, He I at 667.8 nm, and Hα at 656.3 nm).

The transitions from He I, Hα, O I, and OH(A–X) exhibit their maximum intensities in the immediate vicinity of the needle tip (z=0) inside the dielectric tube. The presence of Hα, O I, and OH(A–X) suggests the dissociation of water vapor and oxygen impurities either present in the helium supply or desorbed from the dielectric walls [88]. It is reminded here that the μAPPJ is operated (i.e., plasma "ON") for 30 minutes to ensure stable operation before any experiment takes place. Within the microtube, the four species follow a similar trend, only slightly decreasing in intensity as we move towards the nozzle exit (z=7 mm). Upon exiting the microtube, all intensities drop abruptly within the first few millimeters after the nozzle exit, in agreement with similar studies [53]. This much sharper decay rate in air is a direct consequence of strong collisional quenching of the excited states by $O_2$, $N_2$ and $H_2O$ coming from ambient air.

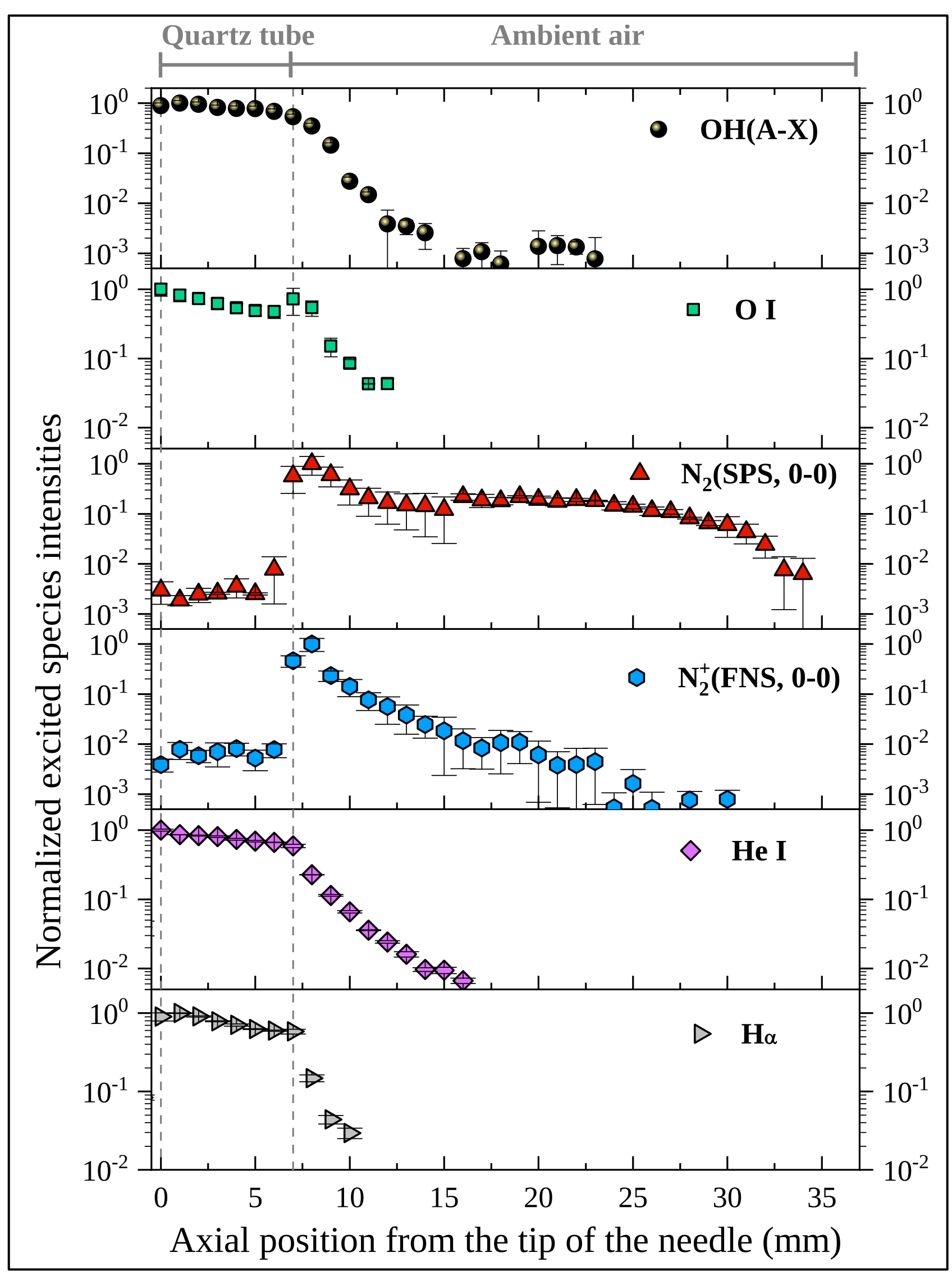

***Figure 13**. Axial evolution of the emission intensity of dominant radiative (excited) species along the discharge propagation axis ($V_P$=9.5 kV, $Q_V$=0.3 slm, $f_{pulse}$=10 kHz, $\Delta t_{pulse}$=160 ns). Different ICCD exposure times are used for each species to maximize the signal-to-noise-ratio (OH(A-X): 0.4 s, O I: 10 s, $N_2$(SPS): 1 s, $N_2^+$(FNS): 0.1 s, He I: 2 s, and Hα: 8 s). The intensities shown in the figure are self-normalized (i.e., normalized to their own peaks). The dashed vertical lines indicate the needle tip (z=0) and the microtube nozzle exit (z=7 mm). Error bars refer to two independent measurements performed on different days.*

Furthermore, the nitrogen-based emissive species ($N_2$(SPS) and $N_2^+$(FNS)) are nearly absent in the microtube. Their strong emissions only start to develop just before the nozzle exit and reach a sharp peak slightly after it, where the effluent interacts with surrounding air. This spatial correlation perfectly aligns with the velocity's peak after the tube exit displayed in **Fig. 12**. The surge in both molecular emissions is driven by the strong interaction of the IW front with the ambient atmosphere (composed mainly of $N_2$ and $O_2$). Particularly, the peak in $N_2^+$(FNS) emission is a hallmark of Penning ionization, where long-lived helium metastables ($He^m$) produced inside the tube collide with nitrogen molecules from surrounding air and ionize them [85,87]. This process enhances the axial electric field strength and provides additional electrons that favor species excitation and boost the IW front velocity for a few millimeters in air, as shown in **Fig. 12**.

Beyond the maximum after nozzle exit, the $N_2^+$(FNS) signal decays continuously, indicating that the high-energy processes required for its excitation become weaker in the mixing layer. However, the $N_2$(SPS) signal exhibits a significantly longer "tail", persisting up to $z \approx 30$ mm. This long-range emission correlates with the yellow-green tail observed in the false-color images (**Fig. 5**) and the stable propagation paths seen in the single-shot images (**Fig. 7**). The persistence of the $N_2$(SPS) emission confirms that at $Q_V$=0.3 slm, the effluent remains sufficiently strong to allow for the continued excitation of nitrogen molecules over a long distance. In contrast, the rapid disappearance of the atomic species (He I, Hα, O I) highlights the efficiency of the ambient air in quenching the primary excited atomic species, effectively shielding the core chemical processes of the jet as it transitions into a nitrogen-dominant effluent.

Summarizing, the results of **Figs. 5–13**, the performance of the μAPPJ is governed by a complex coupling between the electric and the hydrodynamic properties of the effluent. The $Q_V$=0.3 slm/$V_P$=9.5 kV (at $f_{pulse}$=10 kHz and $\Delta t_{pulse}$=160 ns) emerges as an optimal operating regime, providing a well-collimated, laminar helium channel that allows the primary ionization wave front to reach peak velocities of approximately $6\times10^2$ km/s. At this flow rate, the discharge effectively utilizes Penning ionization with ambient nitrogen to sustain a long, focused effluent, while simultaneously suppressing the stochastic branching and turbulent fragmentation observed at higher flow rates.

Having established the structural and dynamic features of the discharge, it is essential to evaluate the thermal impact of the jet on its surroundings. To this end, an estimation of the time-averaged gas temperature ($T_{Gas}$) along the axis of the effluent is performed using three complementary spectroscopic methods. These are based (i) on the best fitting of experimental molecular emission spectra with corresponding synthetic ones (**Figs. 14(a)** and **(b1)**), (ii) on the use of the Boltzmann plot method (**Fig. 14(b2)**), and (iii) on the fitting of pressure-broadened helium lines (**Fig. 14(c)**), as detailed in previous relevant studies (including our works) [7,69,72,73,89–92]. **Figs. 14(a), (b1–2)** and **(c)** illustrate representative fitting examples at the optimal operating regime ($V_P$=9.5 kV/$Q_V$=0.3 slm). The average temperature values shown in **Fig. 14(d)** are determined by performing space-resolved measurements along the effluent (as in **Fig. 13**).

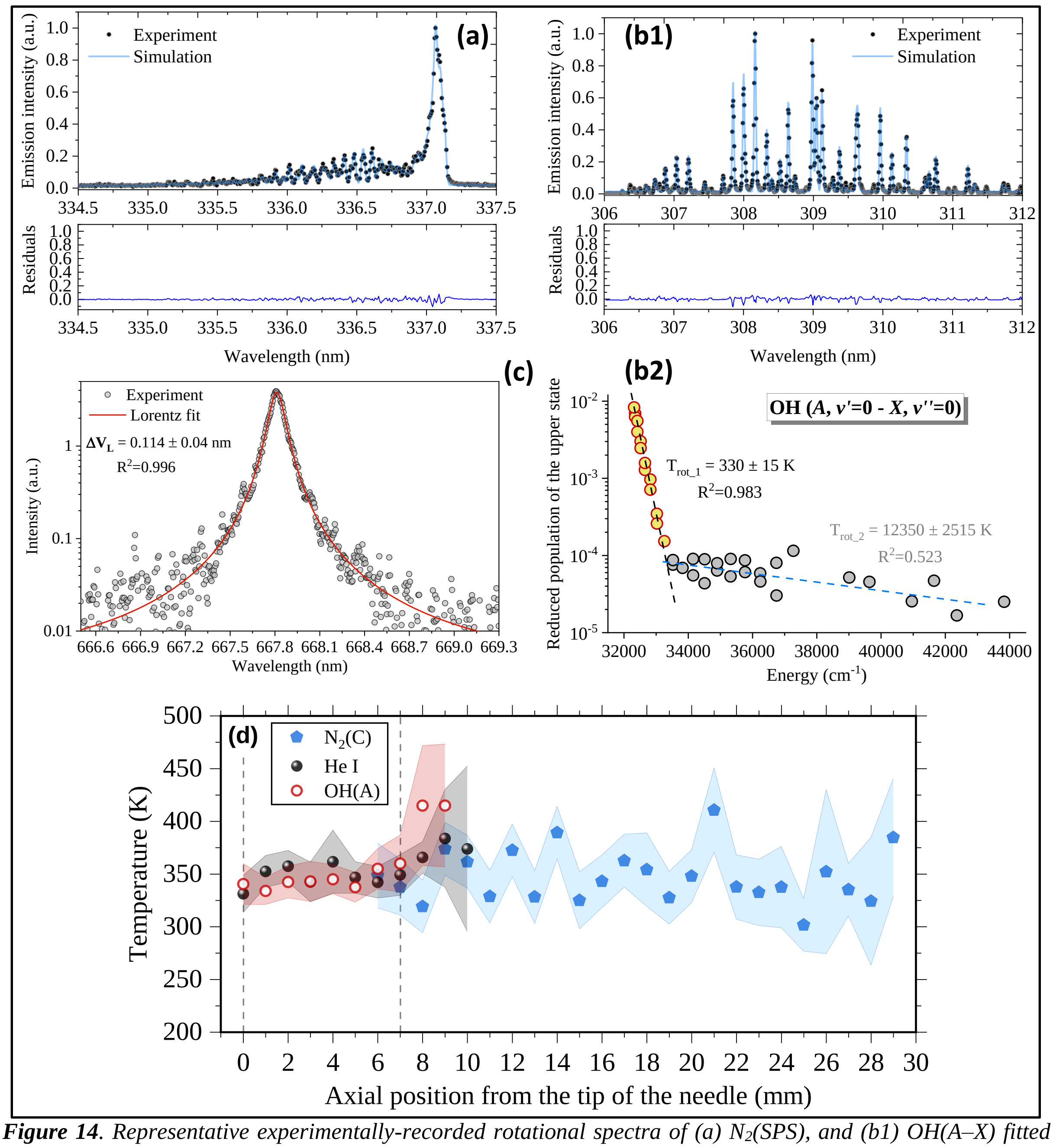

***Figure 14**. Representative experimentally-recorded rotational spectra of (a) $N_2$(SPS), and (b1) OH(A–X) fitted with synthetic corresponding spectra for the determination of the rotational temperatures of $N_2$(C) and OH(A) states, respectively; (b2) Boltzmann plot corresponding to the spectrum in (b1) revealing two rotational temperature groups of the OH(A) state. (c) He I emission line at 667.8 nm fitted with a Lorentzian function to estimate the gas temperature from the broadening of the line. (d) Axial variation of the rotational temperatures of $N_2$(C) and OH(A) and the gas temperature from the He I line (error bars shown in light red, black and blue refer to two sets of experiments performed on different days).*

More specifically, the first method (**Fig. 14(a)**) involves fitting the rotational distribution of the $N_2$(SPS) band ($C^3\Pi_u \rightarrow B^3\Pi_g$, 0–0) with synthetic spectra generated with a custom Matlab® code and cross-compared with massiveOES [14,90–93]. As shown in **Fig. 14(d)**, the best fits yield an average rotational temperature $T_{Rot} \approx 350$ K inside the tube and along the effluent. $N_2$(C) is a common thermometric probe used to estimate $T_{Gas}$ in APPJs [14,72,73]. To further confirm the measured values, the OH(A–X) transition ($A^2\Sigma^+ \rightarrow X^2\Pi$, 0–0) is also recorded (**Fig. 14(b1)**). Hydroxyl radicals typically undergo fast rotational thermalization in atmospheric pressure plasmas, also making them excellent indicators for $T_{Gas}$ [72,89]. However, the OH(A) state may also exhibit a non-Boltzmann distribution, characterized by distinct rotational populations. This is clearly demonstrated in the Boltzamm plot of **Fig. 14(b2)** performed with the massiveOES software [90–92]. A “cold” group of rotationally-excited molecules (red/yellow circles) follows a first linear trend corresponding to an average $T_{Rot}\approx350$ K (**Fig. 14(d)**), in agreement with $N_2$(C), while a “hot” tail of molecules excited at higher rotational levels (black/gray circles) reflects the excess energy from the water dissociation process, as in [89,94]. The temperature of the “cold” group is considered the most reliable estimate of the actual translational gas temperature in our case, as in other similar studies [89,94]. Finally, $T_{Gas}$ is directly obtained through the pressure broadening of the He I emission line at 667.8 nm (**Fig. 14(c)**), a common approach which provides more reliable results in similar operating conditions [69]. By fitting the spectral line profiles with a Lorentzian function and accounting for the pressure-dependent van der Waals and resonance broadening mechanisms, an average gas temperature of about 350 K is obtained between all positions (**Fig. 14(d)**), corroborating the values obtained via the rotational temperatures of $N_2$(C) and OH(A).

Collectively, the three experimental methods used for $T_{Gas}$ estimation in **Fig. 14** are in very good agreement within the experimental uncertainties, placing the average $T_{Gas}$ along the tube and the effluent around 350 K. An increase in the flow rate is expected to induce only a small decrease in $T_{Gas}$ [53]. Thus, the average $T_{Gas}$ remains above room temperature, causing the gas to expand and reduce its density further. Because helium is much lighter than air, we assume that this temperature increase enhances the buoyancy force. The effect of heat on the flow field dynamics is further discussed in **section 5**.

## 5. Discussion

This study presents a multi-modal analysis of a He μAPPJ produced by ns high-voltage pulsing. While earlier research on comparable APPJ systems frequently utilized isolated techniques (such as time- and space-resolved ICCD imaging or OES) occasionally paired with plasma fluid models [50,63,64,66–68], detailed explorations of gas dynamics in μAPPJs remain elusive, while most of the existing studies emphasize on mAPPJ setups [30–33,35–38]. This contribution investigates the electrohydrodynamic characteristics of a ns-pulsed He μAPPJ by uniquely integrating electrical characterization experiments (**Figs. 4**, **8**), space/time-resolved and single-shot ICCD imaging (**Figs. 7**, **9**, **10**), CFD simulations (**Fig. 18**), schlieren photography (**Fig. 6**), and space-resolved OES (**Fig. 13**). The experimental configuration involves helium gas at $T_{Gas}$=293 K flowing through a hollow needle with an inner diameter of 180 μm into a capillary microtube with an inner diameter of 600 μm. Upon application of HV to the needle electrode, a hotter luminous effluent is generated ($T_{Gas}\approx350$ K; **Fig. 14**).

Both schlieren photography (**Fig. 6**) and time-averaged emission data (**Fig. 5**) indicate $Q_V$=0.3 slm as the optimal flow rate for achieving a collimated, elongated, and reproducible effluent. At flow rates below this threshold ($Q_V$<0.3 slm), insufficient flow momentum leads to a slightly bent, "bush-like" appearance, as evidenced in ICCD (**Figs. 5, 7**, and **9**) and schlieren (**Fig. 6**) imaging. In contrast, flow rates exceeding 0.3 slm trigger significant hydrodynamic instabilities (**Fig. 6**), resulting in increased branching (**Fig. 7**) and broader emission patterns (**Figs. 5, 6**, and **9**). Similar hydrodynamic transitions were noted by Foletto *et al.* in a He mAPPJ operating at much higher flow rates (2–7.3 slm), where exceeding a critical threshold caused a shift to turbulent mode [30]. This transition resulted in a broader, conical effluent tip that moved closer to the nozzle, a trend consistent with our observations (**Fig. 6**) despite using a He μAPPJ and significantly lower flow rates (0.1–1 slm). Furthermore, Xian *et al.* employed conventional time-averaged photographs and observed that increasing flow rates induced branching in a pulsed He mAPPJ [33]. They attributed the appearance of branches to the transition of the helium flow from laminar to turbulent regime. This parallels our **Fig. 7** results that, moreover, reveal important details about branch separation, orientation, number and length based on single-shot ICCD photos.

The morphology of the luminous effluent in the present work is sensitive to the voltage amplitude ($V_P$; **Figs. 5**, **7**). Increasing $V_P$ results in faster ionization wave fronts (**Fig. 12**), a finding that aligns with prior research [3,11,18,95]. These IW fronts traverse greater distances along the effluent (**Figs. 7**, **12**), which may amplify minute instabilities at the jet's edge. Wang *et al.* utilized schlieren photography in a mAPPJ to demonstrate that the laminar region length gradually shortens as $V_P$ increases from 5 to 8 kV [35], a range comparable to the present study. This suggests that more "aggressive" ionization waves are more prone to deviating from the central axis, following unstable but locally favorable paths in the mixing zone. Similar bifurcated and distorted averaged emission patterns linked to voltage amplitude have been documented in mAPPJ literature. Shi *et al.* used averaged ICCD imaging to record unstable patterns in sub-µs pulsed mAPPJs at $V_P$>6 kV, with distinct bifurcations appearing at $V_P$>8 kV [95]. They attributed this to shear-layer instabilities caused by intensified interactions between the helium jet and the surrounding air. Additionally, Xian *et al.* observed branched jets at the higher $V_P$ values within the range 3–9 kV when $f_{pulse}$ was set to 4 kHz [33]. Interestingly, this was not observed at 8 kHz, where effluent length increased continuously with voltage, a trend reflected in our time-averaged ICCD images in **Fig. 5** as well. However, our single-shot ICCD measurements (**Fig. 7**) reveal branching at the jet tip starting at $V_P$=6 kV, even at the optimal flow rate of 0.3 slm where averaged images appear stable. This underscores the necessity of combining time-averaged and single-shot ICCD measurements to better capture the complex dynamics of the luminous effluent.

Furthermore, the gas flow rate significantly impacts the sustainability of the ionization wave front velocity (**Figs. 9, 12**), exhibiting a two-phase behavior synchronized with the voltage pulse edges [4,19,20,54,69]. During the rising edge, a high-velocity, solid-core IW front forms. Time-resolved ICCD imaging (**Figs. 9** and **10**) shows the IW front undergoing secondary acceleration at the nozzle exit, with velocities peaking near $6\times10^2$ km/s at 0.3 slm (**Fig. 12(a)**), a behavior that is consistent with other publications [14,51,74,95]. This peak is qualitatively correlated with the OES results (**Fig. 13**), which show a sudden surge in the $N_2$(SPS) and $N_2^+$(FNS) emissions at the exit nozzle, in agreement with previous results [53]. Particularly, the surge in $N_2^+$(FNS) emission strongly suggests the occurrence of Penning ionization, a process in which long-lived helium metastables collide with nitrogen molecules entrained from the air and ionise them [85,87,96,97]. Conversely, during the falling edge of the voltage pulse, a secondary discharge mechanism is triggered. This phase, initiated by the release of surface charges from the dielectric walls, results in a slow-moving, diffuse emission that remains strictly confined to the central axis (**Figs. 9** and **10**) [4,5,17,19,20].

The ionization wave front reaches its peak velocity, reactivity, and propagation distance at $Q_V$=0.3 slm. By preserving a laminar helium channel (**Fig. 6**), the µAPPJ optimizes ionization and gas excitation at the nozzle exit while avoiding the performance-limiting instabilities observed for other flow rates, especially at $Q_V$>0.3 slm.

Although the device exhibits high instantaneous electrical energies **(Figs. 4**, **8)**, which can induce noticeable higher instantaneous gas temperatures inside a single HV pulse [40], the time-averaged gas temperature ($T_{Gas}$) remains approximately 350 K across all spatial positions and diagnostics **(Fig. 14)**. This moderate temperature is attributed to the small pulse width used (160 ns) and effective heat dissipation. The reliability of this thermal profile is confirmed by the close agreement between the $T_{Gas}$ derived from the He I line and the rotational temperatures of $N_2$(C) and OH(A). This approach aligns with similar reported methodologies [14,69,94,98].

The electrohydrodynamic characteristics of the present system remain intricate. By applying calculations similar to those in [38,99–101], the helium velocity within the needle is shown to scale from 70.3 m/s at 0.1 slm to 703 m/s at 1 slm. Upon entering the microtube, the gas undergoes an abrupt expansion, causing velocities to fall to 6.32 m/s and 63.2 m/s, respectively. The resulting Reynolds numbers (Re) range from 104.6 to 1046 in the needle and from 31.4 to 314 in the capillary microtube. With the plasma "ON" the average gas temperature reaches 350 K (**Fig. 14**). Assuming mass flow continuity, i.e., $\rho_1 A u_1 = \rho_2 A u_2$ (where $\rho_1$, $\rho_2$ represent the densities of the He gas at 293 and 350 K, respectively, $A$ is the cross-sectional area of the capillary microtube, and $u_1$, $u_2$ are the corresponding He gas velocities), gas velocity within the system increases by approximately 17% while the Re number drops by about 11%. Comparable velocity increments observed in AC and pulsed mAPPJs suggested that gas heating is a primary source of flow disturbance [38,99–101]. This thermal mechanism was also considered the chief cause of transient vortices in RF-driven jets [37]. However, alternative mechanisms for flow modification in kHz-driven discharges remain a subject of active debate and are examined in the following sections.

It is important to acknowledge that at velocities reaching 703 m/s (Mach 0.7), the flow enters a compressible regime. While our estimation of gas velocity increments utilizes the incompressible continuity equation as a pragmatic approximation for thermal density changes, a more rigorous gas-dynamic treatment would be required to resolve potential shock-cell structures. Such structures could serve as additional seeds for the shear-layer instabilities that trigger the stochastic branching observed at higher flow rates (**Fig. 7**).

Boselli *et al.* utilized time-resolved schlieren imaging to investigate flow modifications in ns-pulsed He mAPPJs [44]. At repetition frequencies ($f_{\text{pulse}}$) below 125 Hz, they observed a turbulent front forming upon plasma ignition and traveling with the gas flow between voltage pulses. However, for $f_{pulse}$>kHz, as in our work, the turbulent front's travel time matched or became higher than the voltage period; consequently, perturbations were not visible between pulses, and turbulent eddies determined the flow profile in the far effluent. The authors suggested mechanisms such as gas heating, localized pressure increases, changes in gas transport properties, and EHD forces. Exploring frequency effects, Wang *et al.* noted that while turbulent fronts appeared at low frequencies (down to 50 Hz), the laminar region shortened as the frequency increased up to 2 kHz [35]. They proposed that higher repetition frequencies lead to the accumulation of thermal effects and energy disturbances, causing more significant flow instabilities than single pulses. Complementing these observations, numerical simulations by Lietz *et al.* on a similar electrode configuration found an average gas temperature of 350 K after a single HV pulse [40], consistent with the results in **Fig. 14**. They reported that high electron density and temperature near the powered electrode caused highly-localized energy deposition, raising the gas temperature by nearly a factor of three within less than 100 ns. This rapid heating generated an acoustic wave that distorted the helium–air shear layer, making it susceptible to Kelvin-Helmholtz instabilities. As a result, shear instabilities develop even after a single HV pulse and propagate with the gas velocity to induce unsteady flow profiles. Hasan *et al.* also simulated thermal effects in a He mAPPJ concluding that the flow field is mainly affected through localized heating, creating pressure gradient forces and increasing the gas velocity at the nozzle exit up to 3 m/s [41]. Other experimental studies further support these findings, also highlighting the role of driving voltage, frequency, and gas flow rate in the onset of plasma-induced flow instabilities [32–35,45,86].

Another quantity correlated with flow field modifications is the EHD force, which involves the transfer of momentum from ions to neutral particles. While some works suggested that this effect is negligible within atmospheric pressure plasma jets [38,99–101], other research presents a different perspective. For instance, Papadopoulos *et al.* utilized dedicated CFD simulations for a kHz-driven AC helium mAPPJ, incorporating the EHD force term directly into the momentum equation to account for discharge effects [31]. Their model showed good agreement with schlieren photographs when using an EHD force amplitude of $5\times10^3$ N/m$^3$, a value deemed sufficient to trigger electrohydrodynamic effects in the effluent. In their study, gas temperature, estimated at approximately 350 K, was not found to be a significant factor in flow pattern alteration. Supporting the idea of combined influences, Whalley *et al.* employed particle image velocimetry (PIV) and schlieren photography on a similar jet, suggesting that turbulent flows arise from the synergy between gas heating and small-amplitude EHD forces [39]. These perturbations destabilize the shear layer at the tube exit and amplify as they move downstream. Further investigations into pulsed helium mAPPJs by Darny *et al.* utilized ICCD and schlieren imaging to examine interactions with metallic targets under varying voltage polarities [42]. They found that plasma generates a force that counteracts buoyancy and enhances helium flow expansion, particularly under negative voltage polarities where long-lived negative ions move in concert with the gas flow. Similarly, van Doremaele *et al.* explored the hydrodynamic traits of pulsed jets impinging on dielectric targets using OES, Rayleigh scattering, and schlieren imaging [102]. In their case, heating and EHD forces from single ionization waves had minimal impact on the flow. Instead, they linked flow modifications to surface charge accumulation, which was more significant during positive pulses. Under these conditions, charges remained on the dielectric long after the voltage pulse, establishing an electric field that drew negative ions toward the target. During negative pulses, however, the motion of negative ions opposed the helium stream while positive ions moved toward the target. Additionally, Iseni *et al.* investigated a similar positive-polarity helium jet, employing OES to measure pressure-broadened He I lines [103]. They calculated EHD force amplitudes of (3–6)×$10^6$ N/$m^3$ near the nozzle; for downward-facing jets, this force counteracted buoyancy, while upward-facing configurations showed a smaller change in line width, suggesting an air fraction of (16±5)% at the jet tip.

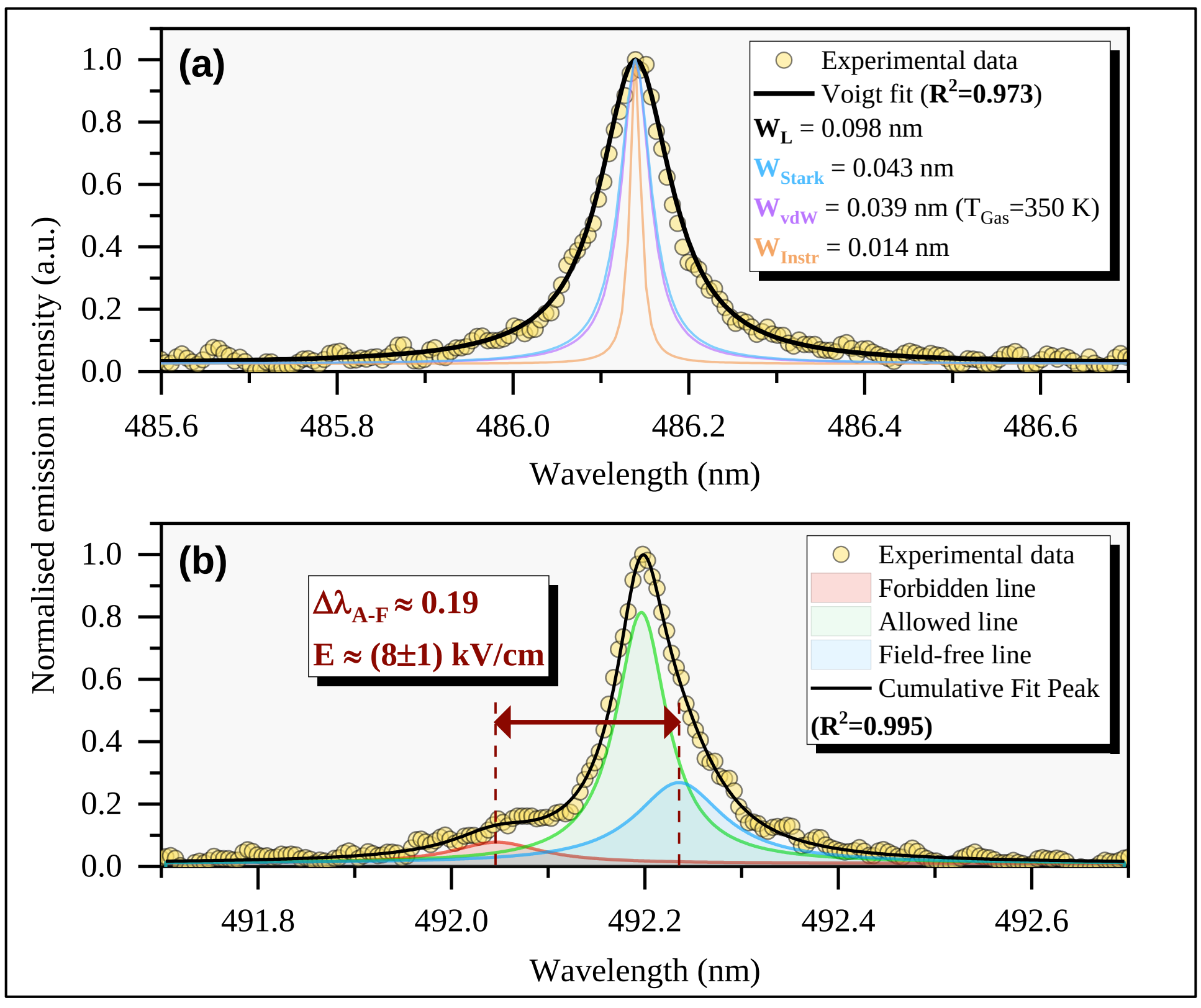

***Figure 15**. (a) Indicative time-averaged (8×$10^4$ HV pulses) Hβ line at the capillary exit fitted with a Voigt function (black line) to extract the electron density. The line profile is dominated by the Lorentzian width ($W_L$) from which the Stark width ($W_{Stark}$; blue line) is deduced by subtracting van der Waals ($W_{vdW}$; purple line) and instrumental ($W_{Instr}$) widths (orange). (b) Indicative π-polarized time-averaged (8×$10^4$ HV pulses) emission spectrum of the He I line at 492.19 nm. $\Delta\lambda_{A\text{-}F}$ denotes the distance between the allowed and forbidden lines used for the calculation of the axial electric field strength.*

In our experiments, the force per unit volume ($f_{PUV}$) can be approximated based on the diagnostic data obtained. The $f_{PUV}$ induced by a single streamer is approximated using the formula [31,104]: $f_{PUV} \approx n_i e E_{loc}$, where $n_i$ is the ion density in the streamer head (here assumed to be equivalent to the electron density, since negative ion density is proposed to be negligible in pure He plasma jets [79]), $e$ is the elementary charge, and $E_{loc}$ is the local electric field. The time-averaged electron density ($\overline{n_e}$) is found to be ~$10^{14}$ $cm^{-3}$ at the tube exit using the Stark broadening of the Hβ line, while the time-averaged axial electric field ($\overline{E_{loc}}$) is estimated to be ~8 kV/cm via Stark polarization spectroscopy (using the He I transition at 492.19 nm). For this analysis, we used standard methodologies detailed in relevant references [12,74–78]. Representative fitting results of the Hβ and He I lines used in this work to obtain $\overline{n_e}$ and $\overline{E_{loc}}$, respectively, are shown in **Fig. 15**.

By substituting the values for $\overline{n_e}$ and $\overline{E_{loc}}$ in the aforementioned formula, the $f_{PUV}$ averaged over the HV pulse duration is approximately $10^7$ $N/m^3$, which is close to the values reported by Iseni *et al.* [103]. This calculation should only be treated as a rough estimation since we assumed that $n_i$=$n_e$. In the case of guided ionisation waves in helium plasma jets, the electric wind is generated in the non-neutral region of the leading front of the ionization wave, where $n_i \neq n_e$ [79]. In principle, $f_{PUV}$ is modulated by the external electric field and the localized field of the ionization wave front [79,104]. The duration of the propagation of the ionization wave corresponds in our µAPPJ system to a time interval of 160 ns (i.e., HV pulse width) (**Fig. 11**). However, our discharge is operated at 10 kHz (i.e., 100 µs period). As proposed by Boeuf *et al.*, the time-averaged EHD force per unit volume ($\overline{f_{PUV}}$) can be determined using the simplified equation [104]:

$$\overline{f_{PUV}} = \varepsilon_0 \frac{V^2}{s^3} F \delta t \tag{5}$$

with $\varepsilon_0$ being the vacuum permittivity, $V$ the potential drop in the cathode sheath (here considered equal to $V_P$=9.5 kV, as per previous works [36]), $s$ the sheath length, $F$ the discharge frequency (here 10 kHz), and $\delta t$ the ratio between the sheath length and the ionization wave propagation velocity. To obtain $\overline{f_{PUV}}$ from eq. (5), the sheath length is required. Previous studies in mAPPJs considered a sheath length between 100–500 µm [36,100,101]. Besides, Boeuf *et al.* used a sheath length $s$=50 µm for a surface air DBD [104]. For this range of sheath lengths and an ionization wave velocity of 600 km/s at the capillary exit ($Q_V$=0.3 slm, z=7 mm; **Fig. 12**) one obtains $\overline{f_{PUV}}$ ≈$5.3\times10^3$ $N/m^3$ ($s$=50 µm) and $\overline{f_{PUV}}$ ≈$5.3\times10^1$ $N/m^3$ ($s$=500 µm). The former calculated value is in good agreement with the work of Papadopoulos *et al.*, who demonstrated its impact on the flow profile of a kHz AC-driven He plasma jet [31]. However, other studies performed in pulsed-driven helium and argon mAPPJs claim that even an $\overline{f_{PUV}}$ ~$10^3$ $N/m^3$ has an insignificant impact on the flow field [36,41,101].

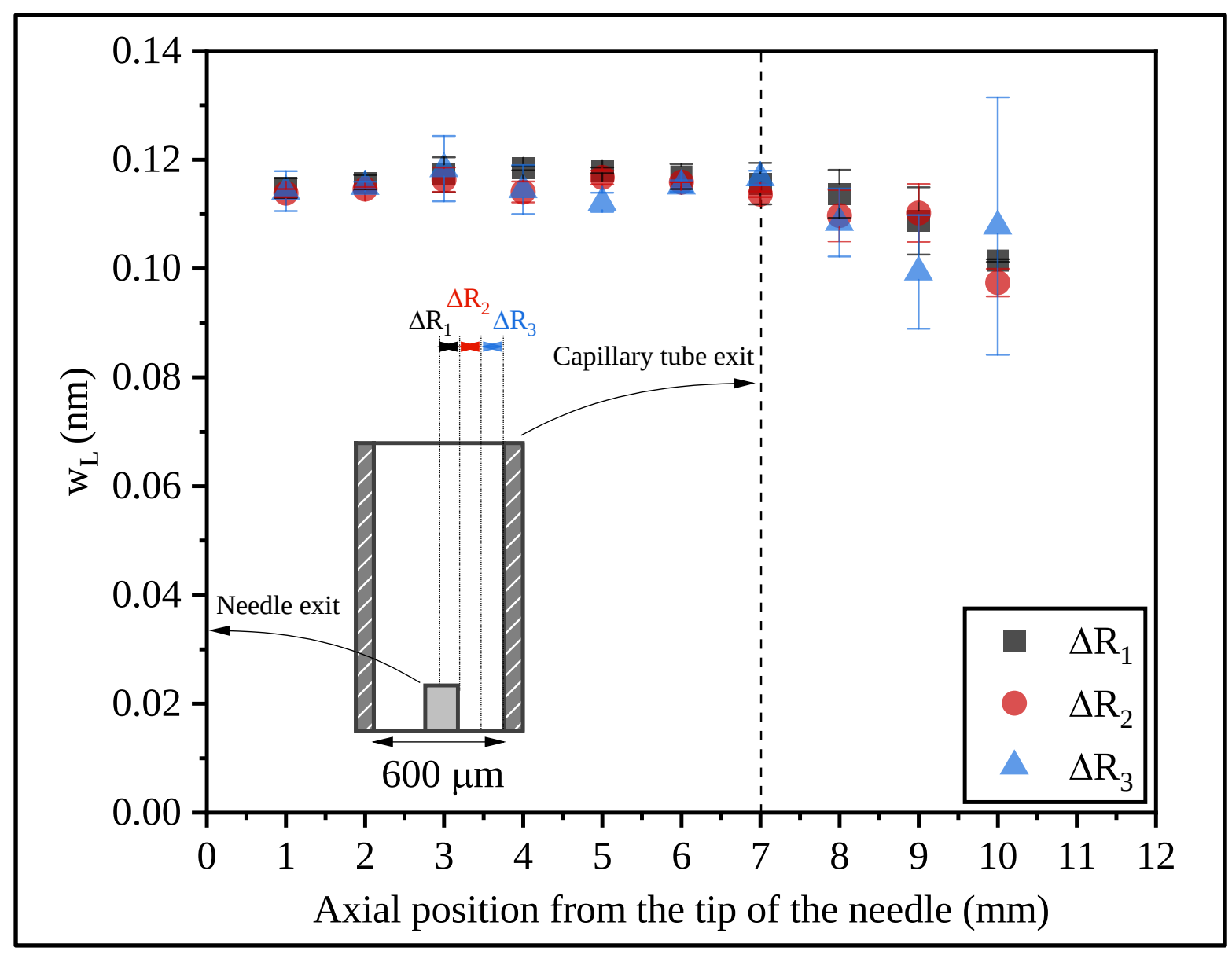

***Figure 16***. *Variation of Lorentzian width of the He I line at 667.8 nm versus the axial position from the tip of the needle for three radial zones ($\Delta R_1$, $\Delta R_2$, $\Delta R_3$; shown in the inset). Each zone has an approximate thickness of 100 μm. Conditions: $V_P$=9.5 kV, $\Delta t_{pulse}$=160 ns, $f_{pulse}$=10 kHz, $Q_V$=0.3 slm. Error bars refer to two series of experiments on different days.*

An alternative to estimate the EHD force gradient can be achieved using the formula [103]:

$$\frac{\Delta p_{He}}{\Delta z} = \frac{1}{A}\frac{\Delta f}{\Delta z} \tag{6}$$

where, $\Delta p_{He}/\Delta z$ is the helium pressure gradient on the effluent axis, z, while $\Delta f/\Delta z$ is the force gradient applied to an area $A$ (corresponding here to the nozzle cross-section). Calculation of $\Delta f/\Delta z$ requires evaluation of the pressure difference, $\Delta p_{He}$, which can be achieved using the pressure broadening of a resonant He I line [103]. Here, the He I line at 667.8 nm is used that is well-fitted with a Lorentzian function (**Fig. 14(c)**). The variation of the Lorentzian width ($W_L$) inside and outside the capillary within three distinct radial zones ($\Delta R_1$, $\Delta R_2$, $\Delta R_3$; approximately 100 μm each) is shown in **Fig. 16**. The results of **Fig. 16** are consistent with those presented by Iseni *et al.* for a similar He I line at 728.1 nm in an upward facing pulsed He mAPPJ [103]. In **Fig. 16**, $W_L$ globally remains stable inside the tube until the exit nozzle (z=7 mm) and then slightly decreases within the first 3 mm along the effluent. The $W_L$ uncertainty progressively increases in the radial direction (i.e., from $\Delta R_1$ to $\Delta R_2$ to $\Delta R_3$), indicating a larger perturbation of the He flow channel from ambient air species at its periphery ($\Delta R_3$) compared to its core ($\Delta R_1$). For $W_L$≈0.115 nm at the tube exit one finds $p_{He}$=2.5 kPa which corresponds to an average force of 2.5×$10^6$ N/m$^3$ being in good agreement with [103] (3×$10^6$ N/m$^3$; 1 mm after the tube exit). These two numbers differ significantly from the indicative value $\overline{f_{PUV}}$ ~$10^3$ N/m$^3$ estimated using eq. (5) and also reported in other works [36,101]. However, as discussed by Boeuf *et al.* [104], the calculated $\overline{f_{PUV}}$ using eq. (5) exhibits a large error margin (~2 orders of magnitude) due to the uncertainty of the sheath propagation velocity. Another source of $\overline{f_{PUV}}$ error in eq. (5) concerns the uncertainty of the sheath length. As discussed previously, reported values in the literature vary up to one order of magnitude.

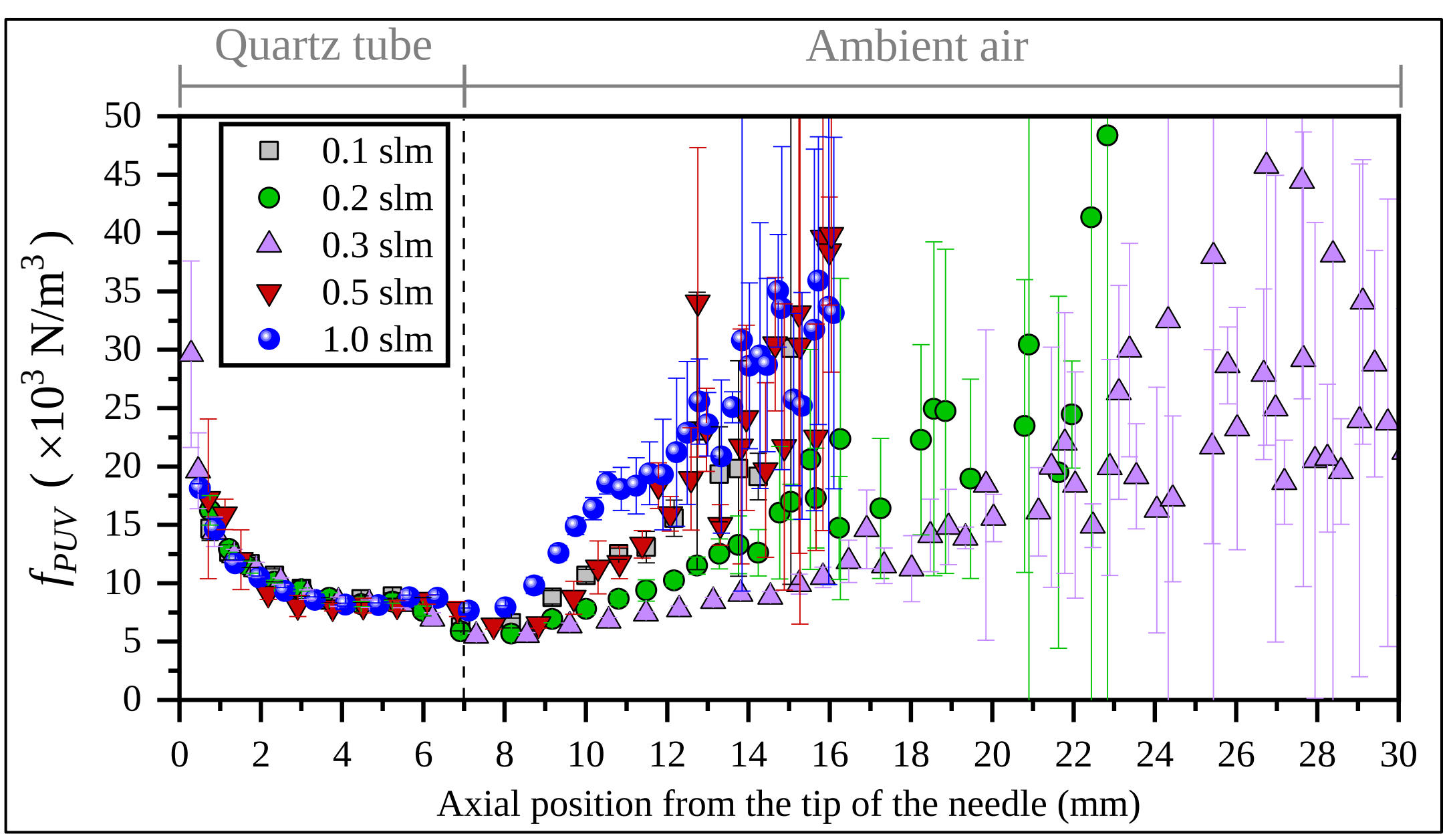

***Figure 17****. Calculated average EHD force ($\overline{f_{PUV}}$) from eq. (5) versus the axial distance from the tip of the needle electrode at indicative gas flow rates.*

In **Fig. 12** the ionization wave velocities along the effluent are higher for $Q_V$=0.3 slm compared to the other flow rates. As per eq. (5), this is expected to give smaller corresponding $\overline{f_{PUV}}$ compared to $Q_V$≠0.3 slm. **Fig. 17** shows the variation of $\overline{f_{PUV}}$ versus the axial position for representative gas flow rates. Here, $\overline{f_{PUV}}$ is derived from eq. (5) using the velocities from **Fig. 12** and a 50 μm sheath length, yielding $\overline{f_{PUV}}$ ~$10^3$–$10^4$ N/m³ being globally consistent with reported values [31,36,101]. While the absolute EHD forces may differ, the relative comparisons of the $\overline{f_{PUV}}$ between different flow rates are meaningful. Inside the tube and within 2 mm from the exit, $\overline{f_{PUV}}$ remains nearly constant for all cases. Further downstream, however, distinct trends appear. At $Q_V$=0.3 slm, $\overline{f_{PUV}}$ increases slowly between z=7 mm and z=20 mm, whereas other flow rates show a sharp rise between z=7 mm and 12 mm. This rapid increase correlates with corresponding unstable regimes, branching, and cloud-like ionization wave fronts observed in **Figs. 6**, **7**, and **9**. These results suggest that an abrupt rise in $\overline{f_{PUV}}$ for $Q_V$≠0.3 slm, possibly in combination with the gas heating (**Fig. 14**), could induce these flow instabilities. Contrary, at 0.3 slm, the quasi-stable $\overline{f_{PUV}}$ in the first 8–10 mm after the nozzle exit and the following moderate increase compared to the other flow rates indicates modest flow perturbation by the plasma, in agreement with **Fig. 6**.

Gas flow modifications in mAPPJs have also been attributed to the presence of residual charges following the passage of an ionization wave front [43,79]. Park *et al.* investigated a horizontally-oriented pulsed He plasma jet combining schlieren photography and ICCD imaging [79]. By varying the voltage pulse width between 1 and 50 μs, they observed that plasma influenced the gas flow and counteracted buoyancy only at longer pulse durations. They concluded that the EHD force generated by a single propagating streamer is negligible, and that the force is primarily induced by residual space charges remaining after streamer propagation. These results were corroborated by the numerical simulations of Papadopoulos *et al.*, who used a dedicated residual charge model to successfully reproduce the flow patterns and plasma structures observed in three different experimental setups [43]. However, in the present work, the HV pulse duration is smaller than 200 ns. According to the previous reports [43,79], this short duration renders the effect of residual charges negligible.

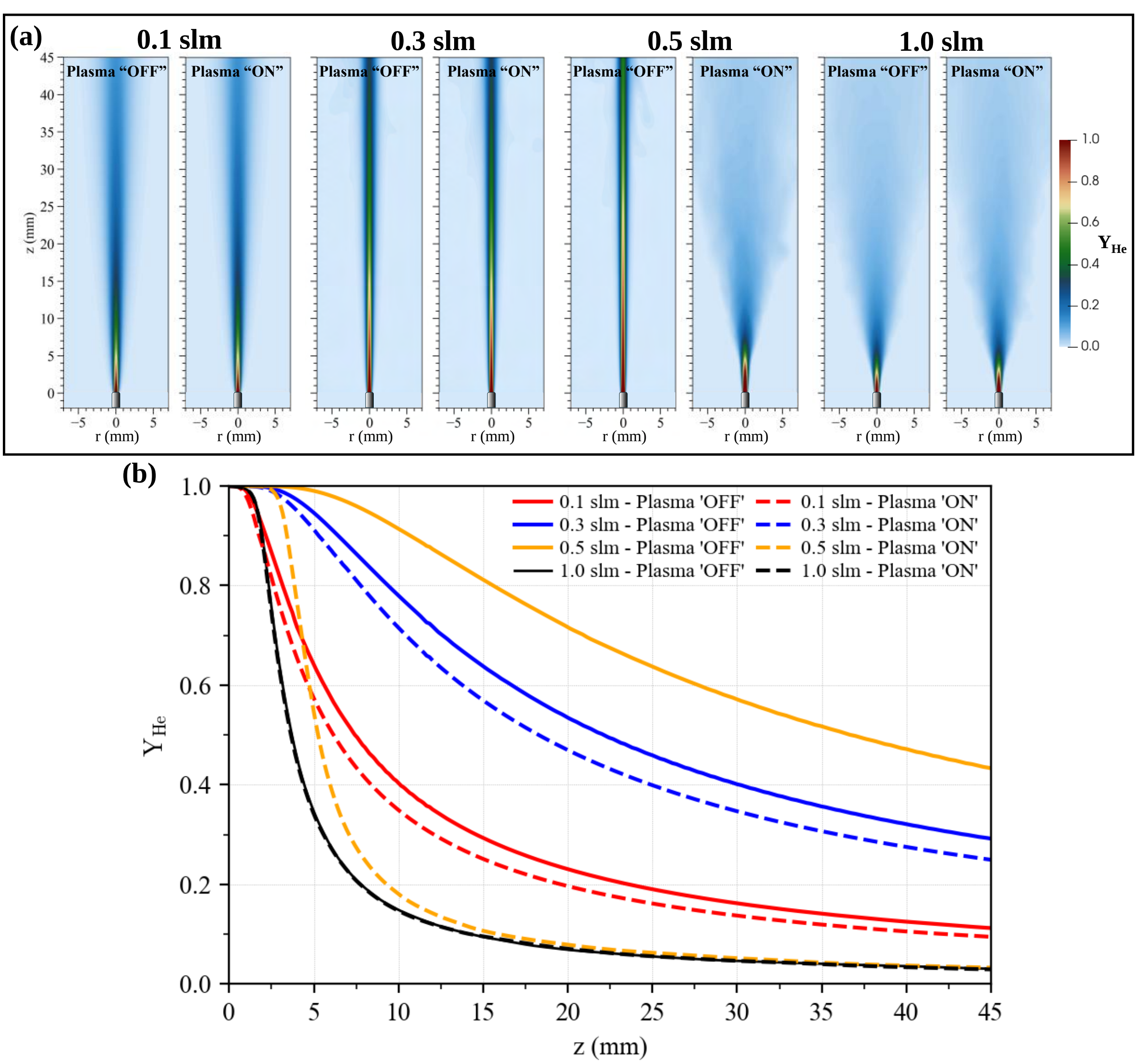

***Figure 18**. (a): Simulated gas flow profiles corresponding to the experimental flow rates shown in Fig. 6 for the following cases: (i) only gas is flown in the capillary tube (i.e., plasma "OFF"; $T_{Gas}$=293 K), and (ii) when plasma is present (i.e., plasma "ON"; $T_{Gas}$=350 K). (b): Helium mass fraction distribution along the effluent centerline (r=0) for plasma "OFF" and plasma "ON" conditions at different flow rates.*

The Lorentzian width values, $W_L$, from **Fig. 16** can be further used to estimate the air fraction ($\chi_{air}$) in the helium channel for $T_{Gas}$=350 K (**Fig. 14(d)**), as follows [105]:

$$W_L^{He\ I\ 667.8\ nm} = (1 - x_{air})\frac{1.79}{T_{Gas}^{0.7}} + x_{air}\frac{3.51}{T_{Gas}^{0.7}} + (1 - x_{air})\frac{26.24}{T_{Gas}} \quad \text{(nm)} \tag{7}$$

As shown in **Fig. 16**, $W_L$ decreases to ≈0.1 nm within the first 3 mm from the tube exit. According to eq. (7), this corresponds to an air mole fraction $\chi_{air}$=10% within the helium channel. Because low signal intensity precluded similar calculations at greater axial distances and different flow rates, we employed CFD simulations to determine the helium mass fraction ($Y_{He}$) along the effluent axis for $Q_V$=0.1–1 slm. These are shown in **Fig. 18(a)** (same flow rates as in **Fig. 6**) for two cases: ambient temperature ($T_{Gas}$=293 K), representing the plasma "OFF" condition, and $T_{Gas}$=350 K, representing the plasma "ON" condition. The simulated profiles are consistent with the gas flow patterns observed in the schlieren images of **Fig 6**, which offer qualitative visualisation but cannot provide $Y_{He}$. The values of $Y_{He}$ extracted from the simulations are depicted in **Fig. 18(b)**.

At $Q_V \leq 0.3$ slm, the flow profiles for plasma “OFF” and plasma “ON” conditions are nearly identical. Plasma ignition induces only a small decrease in $Y_{He}$ along the effluent (10% or smaller), corroborating our previous calculations using eq. (7) and indicating a stable laminar profile. Conversely, at $Q_V$=1 slm, the flow becomes highly unstable and transitions into a turbulent mode. In this regime, the profiles are almost identical regardless of plasma ignition, with $Y_{He}$ dropping sharply to approximately 30% within 5 mm from the tube exit. Furthermore, a notable discrepancy arises at $Q_V$=0.5 slm, where the plasma “OFF” and plasma “ON” profiles differ remarkably. For instance, at z=5 mm, the plasma “ON” flow becomes unstable and adopts an upward-expanding conical shape (see **Figs. 18** and **6**), while $Y_{He}$ falls critically from about 95% to 60%. In this case, plasma ignition strongly perturbs the He channel, leading to significant air penetration into the effluent that disrupts collimation and promotes discharge branching (**Fig. 7**). These observations align with the work of Karakas *et al.*, who identified a transition mode at $Y_{He} \approx 0.45$ characterised by vortex patterns at the effluent tip **[28]**. Their results showed that turbulence leads to a significantly shorter ionization wave front propagation length, which is in good agreement with our experiments (**Figs. 5–9**) and verified by our computer simulations (**Fig. 18**).

Notably, even at the highest flow rate studied (1 slm), the Reynolds number remains below 1100, which, theoretically, suggests a laminar flow regime. However, gas velocities in the needle reach up to Mach 0.7. These high subsonic values make the flow highly energetic, meaning that even minor plasma-induced perturbations, geometric imperfections in the needle or capillary, or slight misalignments can trigger instabilities. Xiong *et al.* investigated these dynamics by correlating the helium mole fraction distribution with different flow rates and capillary diameters **[36]**. By analysing the dimensionless jet length (L/D), i.e., the jet length (L) divided by the capillary tube diameter (D), they demonstrated that the transition to turbulence can occur at much smaller Re numbers than traditionally expected: Re≤200 (laminar), 200<Re≤650 (transitional), and Re>650 (turbulent). L/D values for the effluent lengths reported in **Fig. 8(a)** are plotted against Re in **Fig. 19**. The L/D—Re data distribution closely resembles the findings of Xiong *et al.* for the same capillary diameter. However, our results indicate that the transition regime begins at Re>100, which is half the threshold reported in **[36]** (Re>200). Similar early-onset turbulence has been noted elsewhere; for instance, Foletto *et al.* observed a transition threshold at Re=150 following plasma ignition **[30]**, while Whalley *et al.* **[39]** reported turbulence formation in the effluent after Re=135, which agrees with our observed transition threshold.

While the bulk Reynolds numbers in the capillary microtube (Re=31.4–314) suggest a laminar regime by classical pipe-flow standards, the transition observed at Re>100 (**Fig. 19**) is more accurately described by the instability of a high-speed jet in a confined expansion. In this geometry, the characteristic length for transition is not the pipe wall boundary layer, but the shear layer between the high-velocity helium core and the recirculating peripheral zone. Consequently, Re numbers only serve here as a scaling parameter for jet momentum rather than a metric for fully developed pipe turbulence.

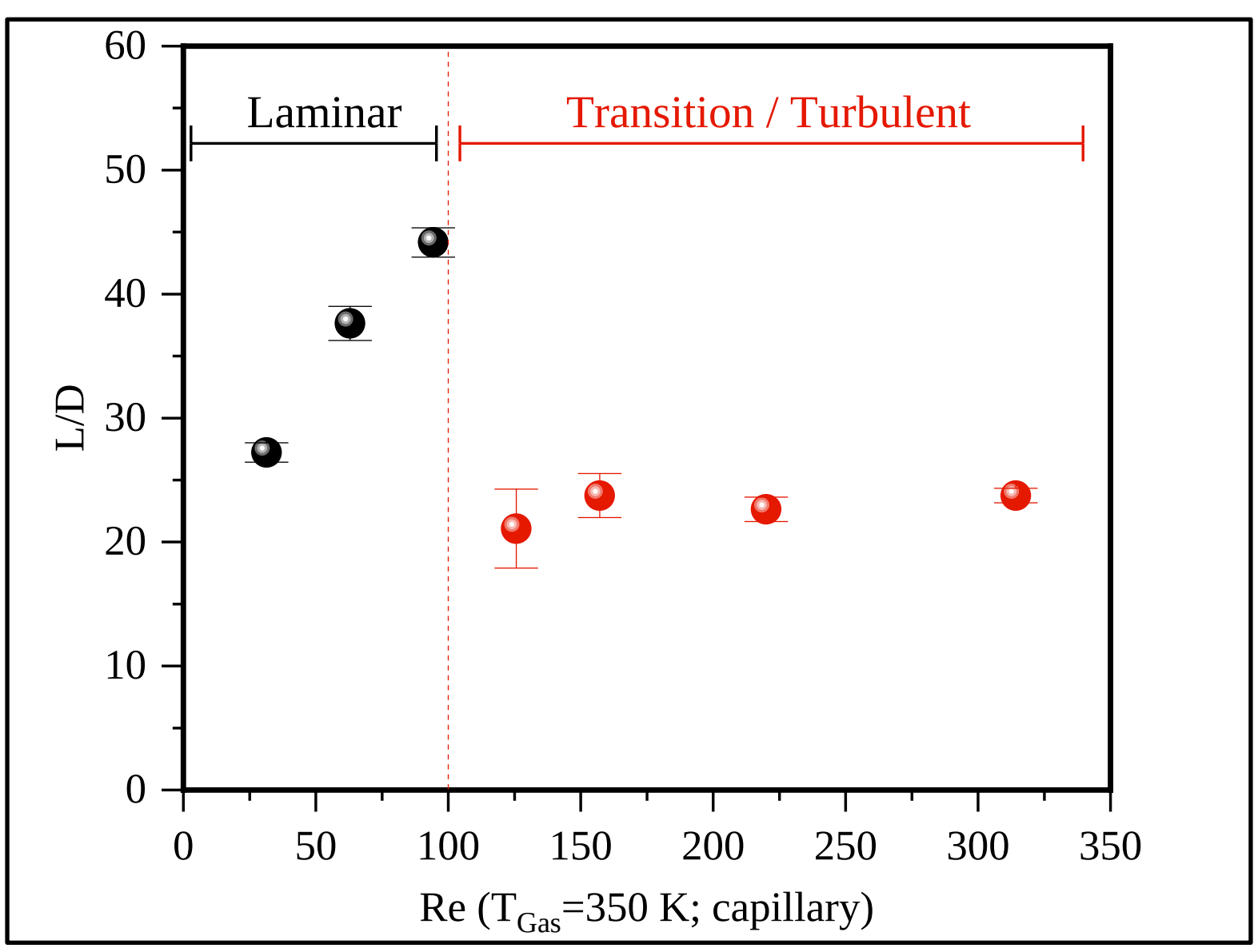

***Figure 19***. *Dimensionless effluent length versus the Re number in the plasma "ON" case.*

For our μAPPJ, the distance between the needle tip and the capillary exit (here $d_3$=7 mm) could affect flow stability, depending on the gas flow rate. This relationship can be roughly evaluated using the hydrodynamic entrance length ($L_E \approx 0.05 \times Re \times d$), where $d$ is the capillary diameter, and the constant 0.05 denotes the point where the centerline velocity reaches 99% of its fully developed value [106,107]. Upon exiting the needle and entering the capillary, the high-velocity gas must transition from a concentrated jet to a parabolic velocity profile as the no-slip condition is established at the capillary walls [7]. Under plasma "OFF" conditions, the calculated $L_E$ values for $Q_V$=0.1, 0.3, 0.5, 0.7, and 1 slm are 0.9, 2.8, 4.7, 6.6, and 9.4 mm, respectively. These drop by about 11% when the plasma is ignited due to the decrease in Re, as mentioned before.

Consequently, for flow rates where $L_E \geq d_3$ (specifically $Q_V$=1 slm), the gas is ejected before the flow becomes fully developed, resulting in inherent hydrodynamic instability. This is consistent with the conical flow profiles observed immediately after the exit of the tube (see **Figs. 6** and **18**). At $Q_V$=0.7 slm, the flow just reaches development before exiting the tube; however, a conical profile still appears (not shown), likely because the transition occurs near the exit where the helium meets ambient air, creating a highly unstable shear layer. Furthermore, geometric defects or a slight needle misalignment can cause $L_E$ to deviate from theoretical values. While flows at $Q_V \leq 0.5$ slm are fully developed at least 2 mm before the exit and remain stable when the plasma is off, they become highly sensitive to localised plasma-induced perturbations when ignited. In the range $0.3 < Q_V \leq 0.5$ slm, these perturbations, combined with potential geometric factors, can alter the flow development and trigger significant instabilities. Here it should be emphasized that the use of the standard laminar entrance length formula provides a simplified first-order scale analysis of flow development, since this assumes a homogeneous plug-flow profile at the inlet. However, given that helium enters the capillary as an energetic jet (reaching up to Mach 0.7), the flow must undergo complex momentum exchange and dissipation within the recirculation regions of the so-called 'backward-facing step' (BFS) or 'sudden expanding flow' geometry. Our μAPPJ is analogous to a BFS geometry, being a standard benchmark case used in fluid mechanics to study flow separation [108]. In a typical BFS flow, a uniform stream from a channel of height $h$ (see **Fig. 20**) encounters a sudden step of height $h$ on one or both walls, resulting in a flow field divided into four distinct regions: the separated shear layer, the recirculation zone, the reattachment region, and the reattached downstream recovery boundary layer. These regions, as they pertain to our setup, are indicatively illustrated in the 2D scheme shown in **Fig. 20**.

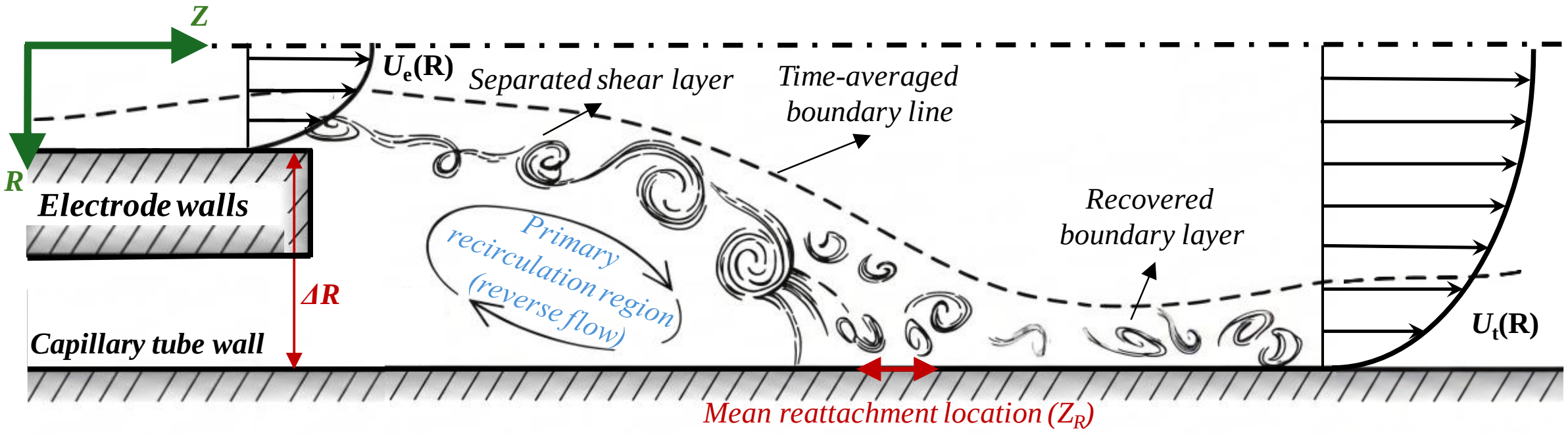

***Figure 20**. Two-dimensional axi-symmetric illustration of a BFS geometry adapted to our μAPPJ configuration (Fig. 1) in the plasma "OFF" case. This illustration is inspired from [108].*

Darny *et al.* investigated a pulsed argon-based mAPPJ with a similar BFS design using time-resolved schlieren photography and space-time resolved ICCD imaging [34]. Their findings revealed that both individual HV pulses and kHz-range HV pulse trains induced flow perturbations; for single pulses, these distortions manifested as turbulent fronts propagating at the flow velocity. Consistent with previous literature [35], the magnitude of these perturbations scaled with pulse frequency and varied with the flow rate. Notably, the BFS geometry facilitates complex internal gas dynamics: the sudden expansion from the needle promotes vortex formation at the internal driven electrode rather than in the external shear layer (**Fig. 20**). The authors proposed that periodic pressure spikes generated near the needle by ionization waves provided mechanical forcing, thereby altering the force balance within the reattachment zone (**Fig. 20**). This forcing can either enhance or suppress the development of large coherent vortices in the recovered boundary layer, depending on the ratio of the forcing frequency to flow rate. This phenomenon persisted into the air and remained independent of the gas type for a constant Re number. Given the structural parallels to our setup, the BFS geometry is clearly a decisive (if not the most important) factor in our effluent stability, governed by the interplay between electrical parameters and gas flow rates. The sudden expansion from the needle promotes vortex shedding at the internal driven electrode. We propose that the ns-pulsed ionization wave fronts provide periodic mechanical forcing that interacts with the reattachment zone. This allows maintaining collimation at 0.3 slm or triggering the fragmentation of the discharge channel for $Q_V$>0.3 slm. For $Q_V$<0.3 slm, the helium jet exhibits a thick, diffuse shear layer (**Fig. 6**). In the frame of BFS context, the low momentum allows the primary recirculation region to dominate early, leading to rapid mixing and air entrainment.

## 6. Conclusions

In this contribution, we investigated the complex interplay between gas dynamics and ns-pulsed discharge properties in a micrometer-scale atmospheric pressure plasma jet (μAPPJ) system operating with helium (He) by combining multiple diagnostics and computational modeling. Experiments performed involve electrical measurements, single-shot and time-resolved ICCD imaging to track the discharge global emission pattern and its dynamic evolution, schlieren imaging to visualize gas dynamics, and space-resolved OES to map the chemical reactivity, gas temperature ($T_{Gas}$), electron density and electric field. Emphasis was given also on identifying the optimal helium flow rate and applied HV amplitude where the effluent reaches enhanced stability and collimation. While gas dynamics of millimeter scale APPJs are widely studied in the literature, the hydrodynamic behavior of μAPPJs remains underreported. Finding the balance between device geometry, electrical parameters, and flow rates in these devices is vital to prevent discharge instabilities for potential applications.

The performance of our He μAPPJ is governed by complex plasma-gas coupling, where the discharge actively modifies its own propagation channel. Our multi-diagnostics approach identified an optimal operating regime at $Q_V$=0.3 slm and $V_P$=9.5 kV ($f_{pulse}$=10 kHz) that enabled low gas consumption and maximized effluent stability and reactivity. At this regime, a balance between momentum-driven collimation and plasma-induced perturbations was reached that resulted in a laminar, elongated luminous effluent reaching a maximum propagation length of approximately 27 mm. Furthermore, this operating regime was observed to limit stochastic branching. Higher HV amplitudes were reported to generate more intense ionization wave fronts that are susceptible to amplifying infinitesimal fluctuations at the helium–air interface. As $V_P$ increased from 4 to 9.5 kV, the ionization wave fronts deflected from the central propagation axis and followed unstable pathways. While branching was noticeably suppressed at 0.3 slm, for $Q_V$<0.3 slm reduced flow momentum allowed buoyancy and ambient air entrainment to become important. This created a "bush-like" morphology of the ionization wave fronts with early-onset branching occurring as close as 7 mm from the nozzle exit. Beyond 0.3 slm, the jet was observed to enter a transitional or turbulent mode that triggered shear-layer instabilities causing the effluent to become wavy, shrunk, and fragmented into multiple stochastic paths. These structures represented a transition where the ionization wave front could no longer follow a single, collimated path and instead produced fragments into multiple stochastic channels. To the best of our knowledge, this is the first study reporting a detailed visualization and quantitative branching analysis using single-shot ICCD images in a He μAPPJ system.

Electrical measurements coupled to time-resolved ICCD imaging revealed a dual-phase discharge mechanism: a high-velocity (up to 600 km/s at optimal conditions) primary ionization wave during the rising voltage edge, and a slower, diffuse secondary discharge during the falling edge. Within the dielectric tube, the primary wave front initiated with an annular structure, concentrating near the walls before evolving into a solid-core ionization wave front upon reaching the nozzle exit. At 0.3 slm, this ionization wave front effectively detached from the main discharge channel to propagate discreetly over the longest distance. This produced a peak velocity immediately after the nozzle exit (z>7 mm) across the spectrum of the flow rates tested (0.1–1 slm). OES performed at the optimal conditions revealed a surge in $N_2^+$(FNS) emission intensity just after the nozzle exit. This indicated that Penning ionization, where long-lived He metastables collide with and ionize nitrogen molecules, as the primary driver for secondary ionization wave front acceleration. The secondary discharge was triggered by the drop of the applied HV amplitude and the release of surface charges (DBD-like behavior) and remained strictly confined in the central axis.

The transition from a stable jet to a turbulent effluent could be driven by the synergy of thermal and electrohydrodynamic forces. Although the estimated time-averaged gas temperature was moderate ($T_{Gas}$≈350 K), localised energy deposition near the powered electrode could cause fast temperature increases. These density gradients could trigger shear-layer instabilities that distort the helium channel, also depending on the flow rate used. Besides, average EHD force approximations (~$10^3$ N/m³) and CFD simulations were performed. These indicated localized plasma-induced perturbations that could be sufficient to trigger shear-layer instabilities, especially in "sudden expansion" geometries like ours. Especially at $Q_V$>0.3 slm, the abrupt rise in these forces after the tube exit could be correlated with important flow instabilities. By utilizing short pulse durations (<200 ns), EHD effects due to residual charges remained negligible in our experiments.

Furthermore, the stability of the proposed μAPPJ system is most probably linked to its complex "backward-facing step" geometry. This seems a decisive (if not the most important) factor in our effluent stability, governed by the interplay between electrical parameters and gas flow rates. The sudden expansion from the needle promotes vortex shedding at the internal driven electrode. We propose that the ns-pulsed ionization wave fronts provide periodic mechanical forcing that interacts with the reattachment zone. This allows maintaining collimation at 0.3 slm or triggering the fragmentation of the discharge channel for $Q_V$>0.3 slm. For $Q_V$<0.3 slm, the helium jet exhibits a thick, diffuse shear layer. In the frame of BFS context, the low flow momentum allows the primary recirculation region to dominate early, leading to rapid mixing and air entrainment.

Moreover, we reported that the transition from a stable collimated jet to a stochastic branched effluent in a He μAPPJ system involves fluidic, thermal, electrical and structural governing parameters. Precise control over discharge behavior requires the synchronization of electrical excitation with the fluidic recovery and mixing timescales, which is vital for the development of reproducible and tunable microplasma sources for mass spectrometry and high-precision surface processing.

## Acknowledgements

This work was funded by the ANR ULTRAMAP Project (ANR–22–CE51–0027), the Labex SEAM Project (ANR–10–LABX–0096; ANR–18–IDEX–0001), and the IDF regional Project SESAME DIAGPLAS. This work was partly supported by the German Research Foundation (DFG), project number 535827833, the ONISILOS project (101034403) and the IgnitePLASMA project (101129853). The authors would like to acknowledge Dr. Laurent Invernizzi for his help with the installation of the experimental setups, and Dr.-Ing. Philipp Mattern (INP Greifswald) and Dr. P. K. Papadopoulos for their insightful comments on fluid mechanics.

## References

[1] Lu X, Laroussi M and Puech V 2012 On atmospheric-pressure non-equilibrium plasma jets and plasma bullets *Plasma Sources Sci. Technol.* 21 034005

[2] Kc S K, Ghimire B, Hong S-H, Oh J-S and Szili E J 2025 How to control the plasma jet production of reactive species for medical therapy? A topical review *J. Phys. Appl. Phys.* 58 143006

[3] Mericam-Bourdet N, Laroussi M, Begum A and Karakas E 2009 Experimental investigations of plasma bullets *J. Phys. Appl. Phys.* 42 055207

[4] Xiong Q, Lu X, Ostrikov K, Xiong Z, Xian Y, Zhou F, Zou C, Hu J, Gong W and Jiang Z 2009 Length control of He atmospheric plasma jet plumes: Effects of discharge parameters and ambient air *Phys. Plasmas* 16 043505

[5] Xiong Q, Lu X, Liu J, Xian Y, Xiong Z, Zou F, Zou C, Gong W, Hu J, Chen K, Pei X, Jiang Z and Pan Y 2009 Temporal and spatial resolved optical emission behaviors of a cold atmospheric pressure plasma jet *J. Appl. Phys.* 106 083302

[6] Oh J-S, Walsh J L and Bradley J W 2012 Plasma bullet current measurements in a free-stream helium capillary jet *Plasma Sources Sci. Technol.* 21 034020

[7] Gazeli K, Svarnas P, Vafeas P, Papadopoulos P K, Gkelios A and Clément F 2013 Investigation on streamers propagating into a helium jet in air at atmospheric pressure: Electrical and optical emission analysis *J. Appl. Phys.* 114 103304

[8] Gazeli K, Svarnas P, Held B, Marlin L and Clément F 2015 Possibility of controlling the chemical pattern of He and Ar "guided streamers" by means of N2 or O2 additives *J. Appl. Phys.* 117 093302

[9] Athanasopoulos D K, Svarnas P, Liapis C M, Papadopoulos P K, Gazeli K, Giotis K, Vafeas P, Vafakos G P, Giannakakis V and Gerakis A 2023 Combination of ICCD fast imaging and image processing techniques to probe species–specific propagation due to guided ionization waves *Phys. Scr.* 98 055609

[10] Sobota A, Guaitella O, Sretenović G B, Krstić I B, Kovačević V V, Obrusník A, Nguyen Y N, Zajíčková L, Obradović B M and Kuraica M M 2016 Electric field measurements in a kHz-driven He jet—the influence of the gas flow speed *Plasma Sources Sci. Technol.* 25 065026

[11] Sretenović G B, Guaitella O, Sobota A, Krstić I B, Kovačević V V, Obradović B M and Kuraica M M 2017 Electric field measurement in the dielectric tube of helium atmospheric pressure plasma jet *J. Appl. Phys.* 121 123304

[12] Hofmans M and Sobota A 2019 Influence of a target on the electric field profile in a kHz atmospheric pressure plasma jet with the full calculation of the Stark shifts *J. Appl. Phys.* 125 043303

[13] Hofmans M, Viegas P, Rooij O V, Klarenaar B, Guaitella O, Bourdon A and Sobota A 2020 Characterization of a kHz atmospheric pressure plasma jet: comparison of discharge propagation parameters in experiments and simulations without target *Plasma Sources Sci. Technol.* 29 034003

[14] Gazeli K, Svarnas P, Lazarou C, Anastassiou C, Georghiou G E, Papadopoulos P K and Clément F 2020 Physical interpretation of a pulsed atmospheric pressure plasma jet following parametric study of the UV–to–NIR emission *Phys. Plasmas* 27 123503

[15] Teschke M, Kedzierski J, Finantu-Dinu E G, Korzec D and Engemann J 2005 High-speed photographs of a dielectric barrier atmospheric pressure plasma jet *IEEE Trans. Plasma Sci.* 33 310–1

[16] Boeuf J-P, Yang L L and Pitchford L C 2013 Dynamics of a guided streamer (‘plasma bullet’) in a helium jet in air at atmospheric pressure *J. Phys. Appl. Phys.* 46 015201

[17] Karakas E, Akman M A and Laroussi M 2012 The evolution of atmospheric-pressure low-temperature plasma jets: jet current measurements *Plasma Sources Sci. Technol.* 21 034016

[18] Walsh J L, Olszewski P and Bradley J W 2012 The manipulation of atmospheric pressure dielectric barrier plasma jets *Plasma Sources Sci. Technol.* 21 034007

[19] Lu X, Naidis G V, Laroussi M and Ostrikov K 2014 Guided ionization waves: Theory and experiments *Phys. Rep.* 540 123–66

[20] Lu X and Ostrikov K (Ken) 2018 Guided ionization waves: The physics of repeatability *Appl. Phys. Rev.* 5 031102

[21] Yan D, Sherman J H and Keidar M 2017 Cold atmospheric plasma, a novel promising anti-cancer treatment modality *Oncotarget* 8 15977–95

[22] Niu G, Knodel A, Burhenn S, Brandt S and Franzke J 2021 Review: Miniature dielectric barrier discharge (DBD) in analytical atomic spectrometry *Anal. Chim. Acta* 1147 211–39

[23] Gazeli K, Hadjicharalambous M, Ioannou E, Gazeli O, Lazarou C, Anastassiou C, Svarnas P, Vavourakis V and Georghiou G E 2022 Interrogating an *in silico* model to determine helium plasma jet and chemotherapy efficacy against B16F10 melanoma cells *Appl. Phys. Lett.* 120 054101

[24] Rajan A, Boopathy B, Radhakrishnan M, Rao L, Schlüter O K and Tiwari B K 2023 Plasma processing: a sustainable technology in agri-food processing *Sustain. Food Technol.* 1 9–49

[25] Satari Abdullah A, Ahmad M H and Mat Saman N 2025 Atmospheric Pressure Plasma Jet Effectiveness in Food Processing: A Review of Current Trends and Opportunities *IEEE Access* 13 166357–84

[26] Robert E, Akishev Y, Bauzin J-G, Collet G, Dozias S, Ermolaeva S, Bocanegra P E, Grillon C, Hajisharifi K, Hocine A, Maaroufi Z, Pouvesle J-M, Prieur F, Rouillard A, Stancampiano A and Vijayarangan V 2025 State of the Art on the Understanding of Plasma Jet Interaction with Skin Tissue Models *Plasma Med.* 15

[27] Wu Z, Wu K, Wu S, Ji J, Fei C, Liu F and Huang J 2026 Overview of Graphene Oxide Reduction: Focusing on Plasma Strategies *ChemNanoMat* 12 e202500435

[28] Karakas E, Koklu M and Laroussi M 2010 Correlation between helium mole fraction and plasma bullet propagation in low temperature plasma jets *J. Phys. Appl. Phys.* 43 155202

[29] Jiang N, Yang J, He F and Cao Z 2011 Interplay of discharge and gas flow in atmospheric pressure plasma jets *J. Appl. Phys.* 109 093305

[30] Foletto M, Puech V, Fontane J, Joly L and Pitchford L C 2014 Evidence of the Influence of Plasma Jets on a Helium Flow into Open Air *IEEE Trans. Plasma Sci.* 42 2436–7

[31] Papadopoulos P K, Vafeas P, Svarnas P, Gazeli K, Hatzikonstantinou P M, Gkelios A and Clément F 2014 Interpretation of the gas flow field modification induced by guided streamer (‘plasma bullet’) propagation *J. Phys. Appl. Phys.* 47 425203

[32] Robert E, Sarron V, Darny T, Riès D, Dozias S, Fontane J, Joly L and Pouvesle J-M 2014 Rare gas flow structuration in plasma jet experiments *Plasma Sources Sci. Technol.* 23 012003

[33] Xian Y B, Hasnain Qaisrani M, Yue Y F and Lu X P 2016 Discharge effects on gas flow dynamics in a plasma jet *Phys. Plasmas* 23 103509

[34] Darny T, Bauville G, Fleury M, Pasquiers S and Santos Sousa J 2021 Periodic forced flow in a nanosecond pulsed cold atmospheric pressure argon plasma jet *Plasma Sources Sci. Technol.* 30 105021

[35] Wang M, Li P, Han R and Chen X 2025 Study on electroaerodynamic characteristics of helium plasma jet under repetitive nanosecond pulsed discharge excitation *J. Phys. Appl. Phys.* 58 135209

[36] Xiong R, Xiong Q, Nikiforov A Yu, Vanraes P and Leys C 2012 Influence of helium mole fraction distribution on the properties of cold atmospheric pressure helium plasma jets *J. Appl. Phys.* 112 033305

[37] Zhang S, Sobota A, Van Veldhuizen E M and Bruggeman P J 2015 Gas flow characteristics of a time modulated APPJ: the effect of gas heating on flow dynamics *J. Phys. Appl. Phys.* 48 015203

[38] Zheng Y, Wang L, Ning W and Jia S 2016 Schlieren imaging investigation of the hydrodynamics of atmospheric helium plasma jets *J. Appl. Phys.* 119 123301

[39] Whalley R D and Walsh J L 2016 Turbulent jet flow generated downstream of a low temperature dielectric barrier atmospheric pressure plasma device *Sci. Rep.* 6 31756

[40] Lietz A M, Johnsen E and Kushner M J 2017 Plasma-induced flow instabilities in atmospheric pressure plasma jets *Appl. Phys. Lett.* 111 114101

[41] Hasan M I and Bradley J W 2016 Reassessment of the body forces in a He atmospheric-pressure plasma jet: a modelling study *J. Phys. Appl. Phys.* 49 055203

[42] Darny T, Pouvesle J-M, Fontane J, Joly L, Dozias S and Robert E 2017 Plasma action on helium flow in cold atmospheric pressure plasma jet experiments *Plasma Sources Sci. Technol.* 26 105001

[43] Papadopoulos P K, Athanasopoulos D, Sklias K, Svarnas P, Mourousias N, Vratsinis K and Vafeas P 2019 Generic residual charge based model for the interpretation of the electrohydrodynamic effects in cold atmospheric pressure plasmas *Plasma Sources Sci. Technol.* 28 065005

[44] Boselli M, Colombo V, Ghedini E, Gherardi M, Laurita R, Liguori A, Sanibondi P and Stancampiano A 2014 Schlieren High-Speed Imaging of a Nanosecond Pulsed Atmospheric Pressure Non-equilibrium Plasma Jet *Plasma Chem. Plasma Process.* 34 853–69

[45] Pei X, Ghasemi M, Xu H, Hasnain Q, Wu S, Tu Y and Lu X 2016 Dynamics of the gas flow turbulent front in atmospheric pressure plasma jets *Plasma Sources Sci. Technol.* 25 035013

[46] Jõgi I, Talviste R, Raud J, Piip K and Paris P 2014 The influence of the tube diameter on the properties of an atmospheric pressure He micro-plasma jet *J. Phys. Appl. Phys.* 47 415202

[47] Talviste R, Jõgi I, Raud J and Paris P 2016 The effect of dielectric tube diameter on the propagation velocity of ionization waves in a He atmospheric-pressure micro-plasma jet *J. Phys. Appl. Phys.* 49 195201

[48] Xia Y, Wang W, Liu D, Yan W, Bi Z, Ji L, Niu J and Zhao Y 2018 Propagation of atmospheric-pressure ionization waves along the tapered tube *Phys. Plasmas* 25 023513

[49] Talviste R, Jõgi I, Tätte T, Part M, Raud J and Paris P 2021 Application of Y–ZrO2 microtubes as dielectric barrier material in a He atmospheric pressure micro-plasma jet *SN Appl. Sci.* 3 208

[50] Talviste R, Jõgi I, Raud J and Paris P 2016 Development of Ionization waves in an Atmospheric-Pressure Micro-Plasma Jet *Contrib. Plasma Phys.* 56 134–45

[51] Wu S, Lu X, Yue Y, Dong X and Pei X 2016 Effects of the tube diameter on the propagation of helium plasma plume via electric field measurement *Phys. Plasmas* 23 103506

[52] Jánský J, Le Delliou P, Tholin F, Tardiveau P, Bourdon A and Pasquiers S 2011 Experimental and numerical study of the propagation of a discharge in a capillary tube in air at atmospheric pressure *J. Phys. Appl. Phys.* 44 335201

[53] Müller S, Krähling T, Veza D, Horvatic V, Vadla C and Franzke J 2013 Operation modes of the helium dielectric barrier discharge for soft ionization *Spectrochim. Acta Part B At. Spectrosc.* 85 104–11

[54] Horvatic V, Michels A, Ahlmann N, Jestel G, Vadla C and Franzke J 2015 Time-resolved line emission spectroscopy and the electrical currents in the plasma jet generated by dielectric barrier discharge for soft ionization *Spectrochim. Acta Part B At. Spectrosc.* 113 152–7

[55] Horvatic V, Müller S, Veza D, Vadla C and Franzke J 2014 Atmospheric helium capillary dielectric barrier discharge for soft ionization: broadening of spectral lines, gas temperature and electron number density *J Anal Spectrom* 29 498–505

[56] Brandt S, Schütz A, Klute F D, Kratzer J and Franzke J 2016 Dielectric barrier discharges applied for optical spectrometry *Spectrochim. Acta Part B At. Spectrosc.* 123 6–32

[57] Schütz A, Klute F D, Brandt S, Liedtke S, Jestel G and Franzke J 2016 Tuning Soft Ionization Strength for Organic Mass Spectrometry *Anal. Chem.* 88 5538–41

[58] Brandt S, Klute F D, Schütz A and Franzke J 2017 Dielectric barrier discharges applied for soft ionization and their mechanism *Anal. Chim. Acta* 951 16–31

[59] Lara-Ortega F J, Robles-Molina J, Brandt S, Schütz A, Gilbert-López B, Molina-Díaz A, García-Reyes J F and Franzke J 2018 Use of dielectric barrier discharge ionization to minimize matrix effects and expand coverage in pesticide residue analysis by liquid chromatography-mass spectrometry *Anal. Chim. Acta* 1020 76–85

[60] Mao J, Thompson T P, McClenaghan L A, Duncan R M, Gilmore B F and Shakeel H 2026 A miniaturized and low-cost atmospheric pressure helium plasma jet device with high antimicrobial efficiency *Environ. Technol. Innov.* 42 104852

[61] Horvatic V, Müller S, Veza D, Vadla C and Franzke J 2014 Atmospheric Helium Capillary Dielectric Barrier Discharge for Soft Ionization: Determination of Atom Number Densities in the Lowest Excited and Metastable States *Anal. Chem.* 86 857–64

[62] Horvatic V, Michels A, Ahlmann N, Jestel G, Veza D, Vadla C and Franzke J 2015 Time- and spatially resolved emission spectroscopy of the dielectric barrier discharge for soft ionization sustained by a quasi-sinusoidal high voltage *Anal. Bioanal. Chem.* 407 6689–96

[63] Horvatic V, Michels A, Ahlmann N, Jestel G, Veza D, Vadla C and Franzke J 2015 Time-resolved spectroscopy of a homogeneous dielectric barrier discharge for soft ionization driven by square wave high voltage *Anal. Bioanal. Chem.* 407 7973–81

[64] Klute F D, Schütz A, Michels A, Vadla C, Veza D, Horvatic V and Franzke J 2016 An experimental study on the influence of trace impurities on ionization of atmospheric noble gas dielectric barrier discharges *The Analyst* 141 5842–8

[65] Klute F D, Michels A, Schütz A, Vadla C, Horvatic V and Franzke J 2016 Capillary Dielectric Barrier Discharge: Transition from Soft Ionization to Dissociative Plasma *Anal. Chem.* 88 4701–5

[66] Tian C, Ahlmann N, Brandt S, Franzke J and Niu G 2021 Optical characterization of miniature flexible micro-tube plasma (FμTP) ionization source: A dielectric guided discharge *Spectrochim. Acta Part B At. Spectrosc.* 181 106222

[67] Gazeli O, Lazarou C, Niu G, Anastassiou C, Georghiou G E and Franzke J 2021 Propagation dynamics of a helium micro-tube plasma: Experiments and numerical modeling *Spectrochim. Acta Part B At. Spectrosc.* 182 106248

[68] Song H, Tian C, Speicher L, Ahlmann N, Brandt S, Niu G and Franzke J 2024 Elucidation of discharge mechanisms in He- and Ar-flexible μ-tube plasmas by temporally and spatially resolved plasma optical emission phoresy's spectroscopy *Spectrochim. Acta Part B At. Spectrosc.* 219 107014

[69] Stefas D, Agha Y, Invernizzi L, Sousa J S, Prasanna S, Franzke J, Anastassiou C, Lombardi G and Gazeli K 2025 Ultrafast ps–TALIF and streak camera diagnostics of atomic hydrogen in a helium microplasma jet *J. Phys. Appl. Phys.* 58 495201

[70] Traldi E, Boselli M, Simoncelli E, Stancampiano A, Gherardi M, Colombo V and Settles G S 2018 Schlieren imaging: a powerful tool for atmospheric plasma diagnostic *EPJ Tech. Instrum.* 5 4

[71] Giotis K, Stefas D, Agha Y, Höft H, Duten X, Svarnas P, Lombardi G and Gazeli K 2025 Discharge dynamics in a cylindrical SDBD prototype reactor under ns-pulsed and sinusoidal AC operation *Phys. Plasmas* 32 113510

[72] Van Doremaele E R W, Kondeti V S S K and Bruggeman P J 2018 Effect of plasma on gas flow and air concentration in the effluent of a pulsed cold atmospheric pressure helium plasma jet *Plasma Sources Sci. Technol.* 27 095006

[73] Iseni S, Michaud R, Lefaucheux P, Sretenović G B, Schulz-von Der Gathen V and Dussart R 2019 On the validity of neutral gas temperature by emission spectroscopy in micro-discharges close to atmospheric pressure *Plasma Sources Sci. Technol.* 28 065003

[74] Britun N, Dennis Christy P R, Gamaleev V, Hsiao S-N and Hori M 2023 Diagnostics of a nanosecond atmospheric plasma jet. Ionization waves, plasma density and electric field dynamics *J. Appl. Phys.* 133 183303

[75] Nikiforov A Y, Leys C, Gonzalez M A and Walsh J L 2015 Electron density measurement in atmospheric pressure plasma jets: Stark broadening of hydrogenated and non-hydrogenated lines *Plasma Sources Sci. Technol.* 24 034001

[76] Remigy A, Kasri S, Darny T, Kabbara H, William L, Bauville G, Gazeli K, Pasquiers S, Santos Sousa J, De Oliveira N, Sadeghi N, Lombardi G and Lazzaroni C 2022 Cross-comparison of diagnostic and 0D modeling of a micro-hollow cathode discharge in the stationary regime in an $Ar/N_2$ gas mixture *J. Phys. Appl. Phys.* 55 105202

[77] Mirzaee M, Simeni Simeni M and Bruggeman P J 2020 Electric field dynamics in an atmospheric pressure helium plasma jet impinging on a substrate *Phys. Plasmas* 27 123505

[78] Iseni S, Nishime T M C and Gerling T 2024 Transition from afterglow to streamer discharge in an atmospheric capacitively coupled micro-plasma jet *Appl. Phys. Lett.* 125 203502

[79] Park S, Cvelbar U, Choe W and Moon S Y 2018 The creation of electric wind due to the electrohydrodynamic force *Nat. Commun.* 9 371

[80] Kelly S, Golda J, Turner M M and Schulz-von Der Gathen V 2015 Gas and heat dynamics of a micro-scaled atmospheric pressure plasma reference jet *J. Phys. Appl. Phys.* 48 444002

[81] Seyfi P, Heidari A, Khademi A, Golghand M, Gharavi M and Ghomi H 2021 The effect of modulated electric field on characteristic of SDBD -like plasma jet for surface modification *Contrib. Plasma Phys.* 61 e202000155

[82] He T, Song L, He Y, Chen Z and Zheng Y 2025 The Effect of Tube Diameter on the Surface Distributions of ROS in Model Tissue Treated with a He + O2 Plasma Jet *Plasma Chem. Plasma Process.* 45 239–53

[83] Anon Computational Fluid Dynamics - The Basics with Applications McGraw-Hill Series in Aeronautical and Aerospace Engineering - (Anderson Jr J.D., (1995))

[84] Weller H G, Tabor G, Jasak H and Fureby C 1998 A tensorial approach to computational continuum mechanics using object-oriented techniques *Comput. Phys.* 12 620–31

[85] Douat C, Kacem I, Sadeghi N, Bauville G, Fleury M and Puech V 2016 Space-time resolved density of helium metastable atoms in a nanosecond pulsed plasma jet: influence of high voltage and pulse frequency *J. Phys. Appl. Phys.* 49 285204

[86] Qaisrani M H, Xian Y, Li C, Pei X, Ghasemi M and Lu X 2016 Study on dynamics of the influence exerted by plasma on gas flow field in non-thermal atmospheric pressure plasma jet *Phys. Plasmas* 23 063523

[87] Britun N and Hori M 2025 Enhancing He $2^3$ $S_1$ state production in the burst regime of a nanosecond discharge *Plasma Sources Sci. Technol.* 34 01LT02

[88] Winter J, Wende K, Masur K, Iseni S, Dünnbier M, Hammer M U, Tresp H, Weltmann K-D and Reuter S 2013 Feed gas humidity: a vital parameter affecting a cold atmospheric-pressure plasma jet and plasma-treated human skin cells *J. Phys. Appl. Phys.* 46 295401

[89] Xiong Q, Nikiforov A Y, Lu X P and Leys C 2010 High-speed dispersed photographing of an open-air argon plasma plume by a grating–ICCD camera system *J. Phys. Appl. Phys.* 43 415201

[90] Voráč J, Kusýn L and Synek P 2019 Deducing rotational quantum-state distributions from overlapping molecular spectra *Rev. Sci. Instrum.* 90 123102

[91] Voráč J, Synek P, Potočňáková L, Hnilica J and Kudrle V 2017 Batch processing of overlapping molecular spectra as a tool for spatio-temporal diagnostics of power modulated microwave plasma jet *Plasma Sources Sci. Technol.* 26 025010

[92] Voráč J, Synek P, Procházka V and Hoder T 2017 State-by-state emission spectra fitting for non-equilibrium plasmas: OH spectra of surface barrier discharge at argon/water interface *J. Phys. Appl. Phys.* 50 294002

[93] Cardoso R P, Belmonte T, Keravec P, Kosior F and Henrion G 2007 Influence of impurities on the temperature of an atmospheric helium plasma in microwave resonant cavity *J. Phys. Appl. Phys.* 40 1394–400

[94] Britun N, Dennis Christy P R, Gamaleev V and Hori M 2022 Diagnostics of a nanosecond atmospheric plasma jet. Electron and ro-vibrational excitation dynamics *Plasma Sources Sci. Technol.* 31 125012

[95] Shi Y, Li A, Huang X, Ning W, Jia S and Shen S 2026 Multi-parameter control of nanosecond-pulsed atmospheric pressure plasma jets: Experiments, turning points, and control guidelines *Phys. Plasmas* 33 013509

[96] Lazarou C, Anastassiou C, Topala I, Chiper A S, Mihaila I, Pohoata V and Georghiou G E 2023 The effect of Penning ionization reactions on the evolution of He with $O_2$ admixtures plasma jets *J. Phys. Appl. Phys.* 56 065203

[97] Li Q, Zhu X-M, Li J-T and Pu Y-K 2010 Role of metastable atoms in the propagation of atmospheric pressure dielectric barrier discharge jets *J. Appl. Phys.* 107 043304

[98] Iseni S, Baitukha A, Bonifaci N, Pichard C and Khacef A 2020 Electrodeless atmospheric secondary induced ionization jet (EASII-jet): Dynamics and properties of a transferred helium plasma source *Phys. Plasmas* 27 123504

[99] Oh J-S, Olabanji O T, Hale C, Mariani R, Kontis K and Bradley J W 2011 Imaging gas and plasma interactions in the surface-chemical modification of polymers using micro-plasma jets *J. Phys. Appl. Phys.* 44 155206

[100] Ghasemi M, Olszewski P, Bradley J W and Walsh J L 2013 Interaction of multiple plasma plumes in an atmospheric pressure plasma jet array *J. Phys. Appl. Phys.* 46 052001

[101] Wang R, Sun H, Zhu W, Zhang C, Zhang S and Shao T 2017 Uniformity optimization and dynamic studies of plasma jet array interaction in argon *Phys. Plasmas* 24 093507

[102] Van Doremaele E R W, Kondeti V S S K and Bruggeman P J 2018 Effect of plasma on gas flow and air concentration in the effluent of a pulsed cold atmospheric pressure helium plasma jet *Plasma Sources Sci. Technol.* 27 095006

[103] Iseni S, Pichard C and Khacef A 2019 Monitoring hydrodynamic effects in helium atmospheric pressure plasma jet by resonance broadening emission line *Appl. Phys. Lett.* 115 034102

[104] Boeuf J P, Lagmich Y, Unfer T, Callegari T and Pitchford L C 2007 Electrohydrodynamic force in dielectric barrier discharge plasma actuators *J. Phys. Appl. Phys.* 40 652–62

[105] García M C, Yubero C and Rodero A 2022 Gas temperature and air fraction diagnosis of helium cold atmospheric plasmas by means of atomic emission lines *Spectrochim. Acta Part B At. Spectrosc.* 193 106437

[106] Fox R W, McDonald A T and Mitchell J W *Fox and McDonald's introduction to fluid mechanics* (Hoboken: Wiley)

[107] Bergman T L and Incropera F P 2011 *Fundamentals of heat and mass transfer* (Hoboken, NJ: Wiley)

[108] Chen L, Asai K, Nonomura T, Xi G and Liu T 2018 A review of Backward-Facing Step (BFS) flow mechanisms, heat transfer and control *Therm. Sci. Eng. Prog.* 6 194–216